\documentclass[iop,apj,tighten]{emulateapj}
\usepackage{apjfonts} 
\usepackage{amsmath,amstext,mathtools}
\usepackage[breaklinks,colorlinks,citecolor=blue,linkcolor=magenta]{hyperref} 
\usepackage[all]{hypcap} 
\usepackage{relsize}
\usepackage{xcolor}

\def\blue#1 {\textcolor{blue}{#1}\ }   
\def\purple#1 {\textcolor{purple}{#1}\ }

\newcommand{\dott}[1]{\skew{3.0}\dot{#1}}
\newcommand{\widebar}[1]{\skew{3.0}\overline{#1}}

\shorttitle{Eccentric Circumbinary Disks}
\shortauthors{Mu\~noz \& Lithwick}

\begin{document}

\title{Long-Lived Eccentric Modes in Circumbinary Disks}
\author{Diego J. Mu\~noz
and
Yoram Lithwick}
\affil{
Center for Interdisciplinary Exploration and Research in Astrophysics, Physics and Astronomy,
Northwestern University, Evanston, IL 60208, USA\\
}

\begin{abstract}
Hydrodynamical simulations 
 show that circumbinary
disks become eccentric, even when the binary is circular. 
Here we demonstrate that, in steady state, the disk's eccentricity behaves as a long-lived free mode 
trapped by  turning points that naturally arise from a continuously truncated density profile.
Consequently, both the disk's precession rate and eccentricity profile
may be calculated via the simple linear theory for perturbed pressure-supported disks.  By formulating
and solving the linear theory we find that (i)  surprisingly, 
 the precession rate is roughly determined by the binary's quadrupole, even when the
quadrupole is very weak relative to pressure; (ii)
  the eccentricity profile is largest near the inner edge of the disk, and falls
 exponentially outwards; and (iii) the results from linear theory indeed agree with what
 is found in simulations.
 Understanding the development of eccentric modes in circumbinary disks is a crucial first step for understanding 
 the long term (secular) exchange of eccentricity, angular momentum and mass between the binary and the gas.
Potential applications include the search for a characteristic kinematic signature in disks around candidate binaries and
 precession-induced modulation of accretion over long timescales.
\end{abstract}


\maketitle



\section{Introduction}
Gas eccentricity is expected to grow in disks inside and around binaries \citep[e.g.][]{lub91,whi94,paa08,kle08}.
In particular, two-dimensional hydrodynamical simulations of circumbinary disks (CBDs) have consistently shown significant eccentricities  ($\sim0.3$),
even when the binary is circular  \citep{mac08, mir17,thu17}.
Typically, the eccentricity is large
near the circumbinary ``cavity'' and steeply declines outward.  This overall behavior has
been seen also in three-dimensional magneto-hydrodynamics simulations \citep{shi12}, which suggests that the  growth
of eccentricity is a robust property of disks around accreting binaries.

The mechanisms for eccentricity growth are thought to be either resonant excitation \citep{lub91,ogi07} or
periodic ``pumping'' via oblique spiral shocks \citep{shi12}. The mechanisms for eccentricity damping are less constrained, although
viscous damping \citep{goo06}, orbit-crossing \citep{ogi01b,sta01}, or combinations thereof are the likely processes behind eccentricity saturation.
The nearly steady eccentricity profiles of some numerical simulations \citep{mir17} suggest that damping and excitation are in near-equilibrium; if that
is the case, eccentricity profiles can, in principle, be sustained through the lifetime of the disk.
 
In the past, the  true longevity of eccentricity in simulations has been difficult to ascertain.
CBD simulations are subject to slowly-evolving transients that die out
on the local viscous timescale.
After that, and provided a constant mass supply is available, 
a steady state can be achieved \citep{mir17,mun19,moo19}, in which the CBD reaches a (quasi) stationary
density profile \citep[see also][]{dem20a}. 
The transient phase exhibits a behavior that is not
representative of that of the vastly longer disk lifetime, and 
it has been argued that many of the accepted outcomes of circumbinary accretion, such as binary migration being inward,
have been inferred from simulations in the transient state \citep{mun20a}. Getting past this transient phase
is crucial for unveiling the true, long-term eccentricity profile of CBDs, since eccentric modes are sensitive
to the background density profile \citep{tey16,lee19b}.

Eccentricity longevity can be verified by comparing simulations to theoretical expectations.
If disk eccentricities truly correspond to long-lived modes, and endure over a large number of disk dynamical times, 
they will have a lasting impact on the binary-disk coupling, as the distributions of torques is bound to be
different from that of circular disks, which is the standard assumption of the theory \citep{gol80,art94}. Additional implications
of long-lived eccentricities include the modification the processes taking place within the disk, such as planet formation \citep[e.g.][]{sil15},
and the use of kinematic observational signatures \citep[e.g.,][]{rega11} as an independent diagnosis tool for disk structure.

In this work, we demonstrate that the CBD eccentricities seen
in hydrodynamical simulations are consistent with long-lived normal modes. Through a comparison of linear analysis to hydrodynamical
simulations in steady-state, we confirm the agreement of the eigenfrequencies and eigenfunctions, finding that
the circumbinary cavity size is crucial in determining the spatial extent and the precession rate of eccentricity profiles.



\section{Linear Theory of Circumbinary Eccentricities}\label{sec:linear}
In simulations of CBDs, an initially circular  disk becomes eccentric. Within a  rather 
short time ($\lesssim1$ viscous time at the CBD
inner edge), the eccentricity profile saturates, and thereafter precesses uniformly \citep{mir17,thu17}.
The questions of how the eccentricity is excited and then saturates are difficult ones \citep[e.g.][]{tey16}.
We hypothesize that the disk's behavior in its saturated state  is a normal mode of the disk, and does
not depend on how the eccentricity is excited or saturated. Such a hypothesis is reasonable, provided  the mode's precession rate is fast
 compared to excitation/saturation.
In what follows, we calculate the disk's 
 eccentricity profile and precession rate in steady state. However, we do not address
the {\it amplitude} of the mode, which  requires one to consider excitation/saturation.

\subsection{Basics of Linear Theory}
The evolution of the complex eccentricity 
$E=e{\rm e}^{{\rm i}\varomega}$,
in a locally isothermal 2D disk of density profile $\Sigma$
and sound speed profile $c_s$, is
governed by
\begin{equation}\label{eq:eccentricity_equation}
\begin{split}
2\Sigma R^2\Omega\frac{\partial E}{\partial t}
=&
\frac{{\rm i}}{R}\frac{\partial}{\partial R}\left(\Sigma c_s^2R^3\frac{\partial E}{\partial R}\right)
+{\rm i}R\frac{d}{dR}\left(\Sigma c_s^2\right)E\\
& -\frac{{\rm i}}{R}\frac{\partial}{\partial R}\left(\Sigma \frac{dc_s^2}{dR}R^3 E\right)
+2\Sigma R^2\Omega\frac{\partial E}{\partial t}\bigg|_{\rm grav}~,
\end{split}
\end{equation}
 \citep{goo06,tey16,lee19a}, with $\Omega(R)$ being the local orbital frequency, and
where we have deliberately omitted terms due to
excitation and damping (see Introduction).
The first three terms on the right-hand side of Equation~(\ref{eq:eccentricity_equation})
are due to pressure (the third one being
a consequence of a radially varying sound speed; \citealp{tey16}). 
 The last term
is due to a non-Keplerian
external potential \citep{goo06}. This term can be derived from
the disturbing function \citep[e.g.,][]{mar13}, after ignoring high-frequency terms
(i.e., under the secular approximation). For a circular binary of semi-major axis $a_{\rm b}$
 and mass ratio $q_{\rm b}=m_2/m_1$, we have,
to linear order in $E$ and second order in $a_{\rm b}/R$, 
\begin{equation}\label{eq:eccentricity_quadrupole}
\frac{\partial E}{\partial t}\bigg|_{\rm grav}
={\rm i}\Omega
f_0(R)E~,
\end{equation}
where
\begin{equation}
f_0 = \frac{3}{4}\frac{q_{\rm b}}{(1+q_{\rm b})^2}
\left(\frac{a_{\rm b}}{R}\right)^2 ~~.
\end{equation}
If the binary has a finite eccentricity $e_{\rm b}$, then the
right hand side of
Equation~(\ref{eq:eccentricity_quadrupole}) includes
an additional forcing term ${\rm i}\Omega f_1(R)E_{\rm b}$,
where $ E_{\rm b}=e_{\rm b}{\rm e}^{{\rm i}\varomega_{\rm b}}$
is the binary's complex eccentricity, and $f_1$ is of
third order in $a_{\rm b}/R$. But throughout this paper,  we consider only the case of
circular binaries.

\subsubsection{Disks with Central Cavities}
\begin{figure}
\includegraphics[width=0.48\textwidth]{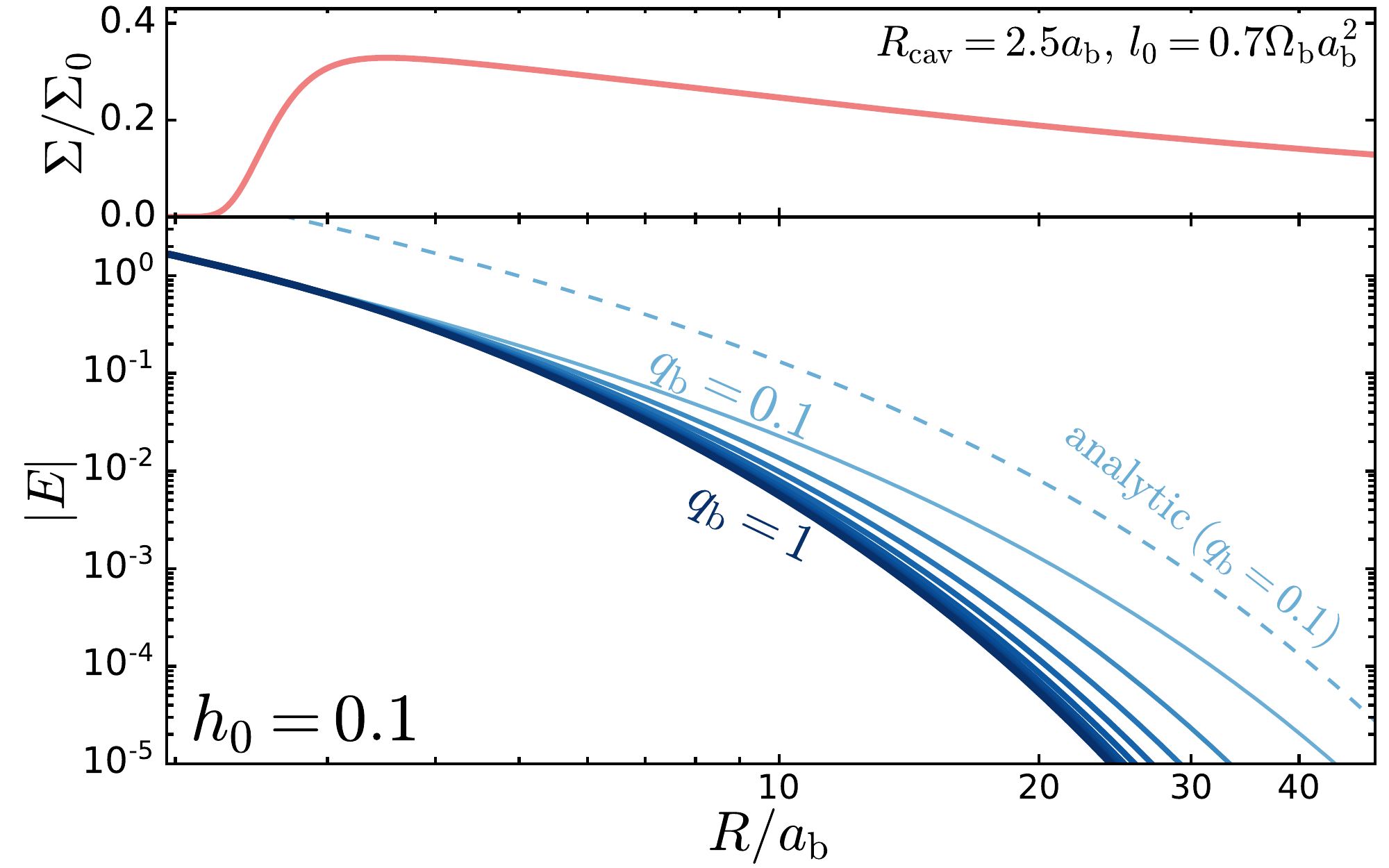}
\vspace{-0.05in}
\caption{
Top: surface density profile (Equation~\ref{eq:density_profile}) for the fiducial parameters
used in this work.
Bottom: numerically computed eccentricity eigenfunctions $|E|$
for different values $q_{\rm b}$ (increasing from lighter to darker blue). 
The dashed curve corresponds to the analytic estimate of $|E|$ (Equation~\ref{eq:approximate_eigenfunction})
evaluated at $q_{\rm b}=0.1$.
\label{fig:eigenfunctions}}
\end{figure}

We adopt a surface density profile representative of a CBD in viscous steady state (VSS)
\citep{mun16b,mir17,mun19}
\begin{equation}\label{eq:density_profile}
\Sigma(R)=
\left[\Sigma_0\left(\frac{R}{a_{\rm b}}\right)^{\!-\frac{1}{2}}\right]
\left[1-\frac{l_0}{\Omega_{\rm b}a_{\rm b}^2}\sqrt{\frac{a_{\rm b}}{R}}\right]
{\rm e}^{-\left(\frac{R_{\rm cav}}{R}\right)^{\!\!\!\!\zeta}}
~.
\end{equation}
The first term in square brackets in Equation~(\ref{eq:density_profile}) is 
the steady-state
solution of a zero-net torque disk with viscosity law $\nu\propto R^{1/2}$ \citep{lyn74}, and approximates
the CBD solution far from the binary. The second term in square brackets is due to an inner ``boundary effect''
\citep{frank2002,pop91}, which modifies the perfect power-law profile whenever the advection of angular momentum by the gas
crossing the disk's inner edge is not exactly balanced by outward viscous transport (see  eq. 19 in \citealp{dem20a}).
The quantity $l_0$, introduced by \citet{mir17}, is the net torque per unit accreted mass exerted by the CBD on the 
binary \citep[see also][]{raf16}, and $\Omega_{\rm b}$ is the binary's orbital frequency.
The final exponential 
factor is the inner cutoff, with two adjustable parameters $\zeta$ and $R_{\rm cav}$.
In addition, we assume $\Omega=\Omega_{\rm b}(R/a_{\rm b})^{-3/2}$,
 and that the disk's aspect ratio ($h_0$) is constant, i.e.,
$c_s^2=h_0^2\Omega_{\rm b}^2 a_{\rm b}^2(R/a_{\rm b})^{-1}$. 
The main adjustable parameters of the problem are $q_b$, $h_0$ and $R_{\rm cav}$.  
The remaining parameters are fix to  $\zeta=12$ (see Section~\ref{sec:hydro_disks} below), and in the present
section, to $l_0=0.7\Omega_{\rm b}a_{\rm b}^2$, which is typical
of simulations \citep{mun20a}. A disk profile with fiducial parameters is depicted in
the top panel of Figure~\ref{fig:eigenfunctions}.

\begin{figure}
\centering
\includegraphics[width=0.45\textwidth]{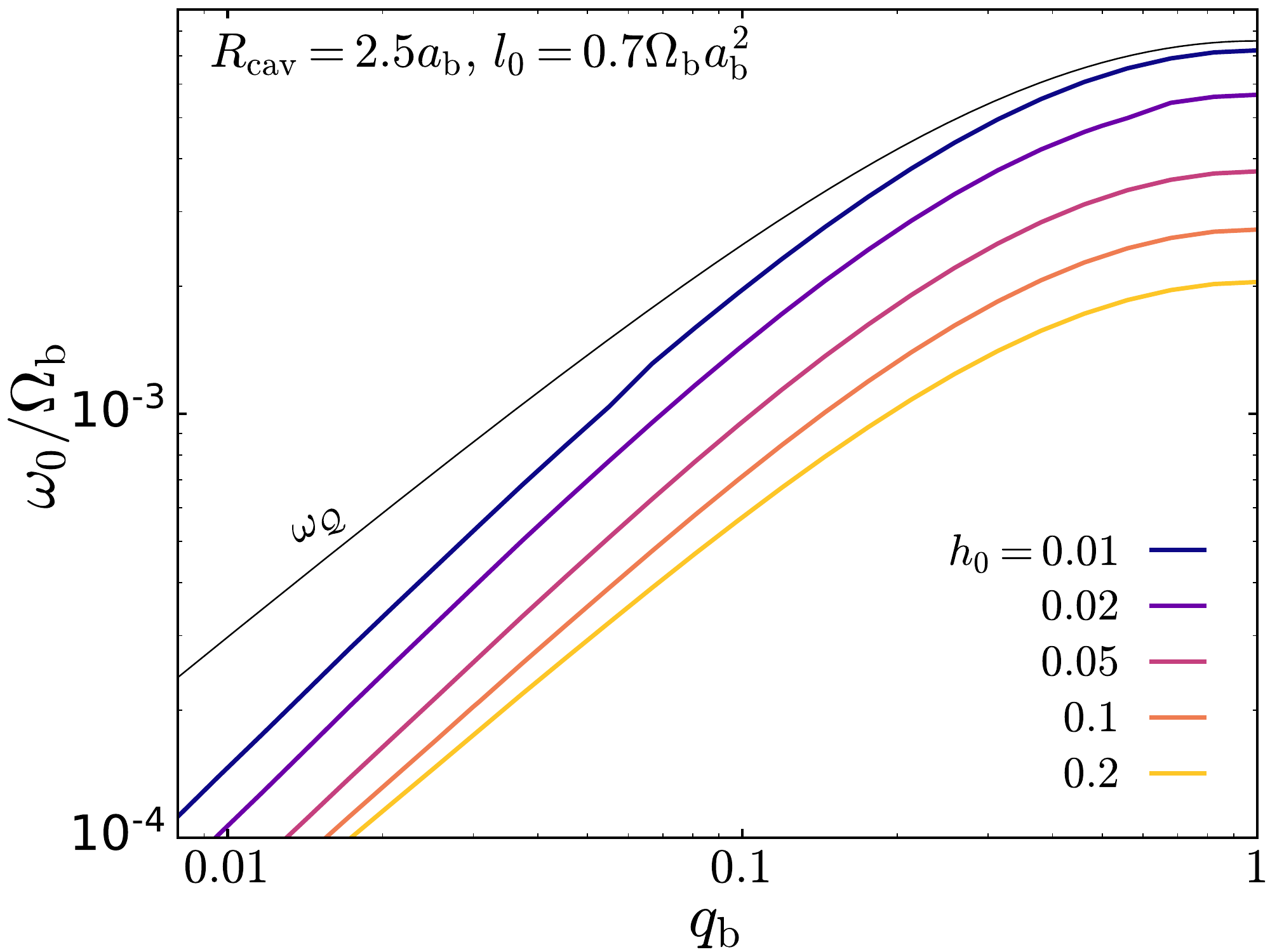}
\vspace{-0.05in}
\caption{Numerically computed eigenfrequency $\omega_0$ (fundamental mode)
in units of $\Omega_{\rm b}$, as a function of $q_{\rm b}$ for different values
of $h_0$.
For comparison, the test-particle precession rate $\omega_{\cal Q}$ (Equation~\ref{eq:characteristic_frequencies}) is shown as thin black line.
\label{fig:eigenfrequencies}}
\end{figure}

\begin{figure}
\centering
\includegraphics[width=0.48\textwidth]{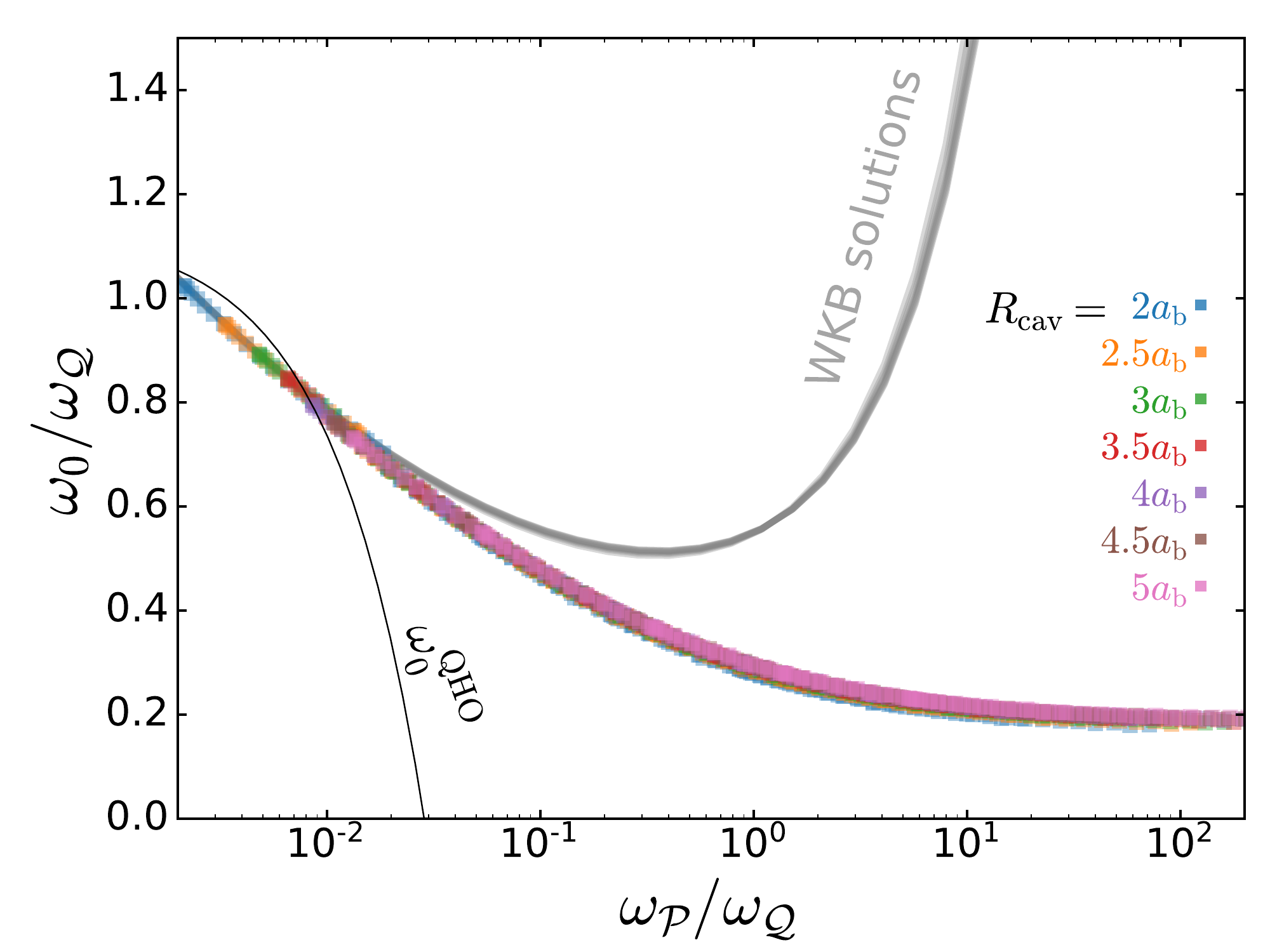}
\vspace{-0.05in}
\caption{Normalized eigenfrequencies $\omega_0/\omega_{\cal Q}$ (color squares) as a function of
 $\omega_{\cal P}/\omega_{\cal Q}$  (Equation~\ref{eq:characteristic_frequencies}),
 for a wide range of values in the parameters
 $h_0$, $q_{\rm b}$ and $R_{\rm cav}$ ($l_0$ fixed). 
Colors denote the values of  $R_{\rm cav}$.
 The approximate analytic
 frequency $\omega_0^{\rm QHO}$ (Equation~\ref{eq:eigenfrequency_approximate}, thin black line)
 and the WKB solutions that satisfy the quantization condition (Equation~\ref{eq:quantum_condition}, thick gray line)
 are shown for comparison.
\label{fig:scaled_eigenfrequencies}}
\end{figure}

\subsection{Numerical Solution of the Boundary Value Problem}\label{sec:linear_numerical}
Inserting solutions of the form $E(t,R)= E(R){\rm e}^{{\rm i}\omega t}$
into  Equation~(\ref{eq:eccentricity_equation})
allows us to replace $\partial E/\partial t$ with ${\rm i}\omega E$.
Together with an appropriate boundary condition 
the eccentricity equation defines a boundary-value problem (BVP) with eigenvalue $\omega$. This BVP can be solved
numerically \citep[e.g.][]{lee19a}, and diverse numerical techniques exist for this purpose \citep[e.g.][]{pryce1993}; 
in this work, we use a shooting method over a domain $R\in[R_{\rm in}, R_{\rm out}]$. At the boundaries, we impose
the boundary condition
$\frac{d}{dR}(E/c_s^2)=0$\footnote{
We require that the Lagrangian pressure perturbation
$\Delta P=\delta P+RE \frac{dP}{dR}$ vanishes at the disk edges,
where $\delta P$ is the Eulerian perturbation \citep{ost67} and $R E$ is the 
radial Lagrangian displacement \citep{pap02}. From the continuity equation, and assuming a locally isothermal equation of state,
$\delta P=-c_s^2R\frac{d}{dR}(E\Sigma)$, and thus
$\Delta P=-R \Sigma c_s^4\frac{d}{dR}(E/c_s^2)$.
}, 
which is appropriate for isothermal disks, instead of the condition ${dE}/{dR}=0$, which is  adequate for adiabatic perturbations. 
In general, the choice of boundary condition has an impact on the BVP eigenfunctions
and eigenvalues, unless the modes are internally trapped in a resonant cavity away from the boundaries (see Section~\ref{sec:wkb} below).
For our particular disk model, the type of boundary condition is irrelevant.
The  computational domain extends
from $R_{\rm in}=0.75 R_{\rm cav}$ to  $R_{\rm out}=450 a_{\rm b}$.
{The location of the inner boundary is chosen to optimize convergence speed, but
only after the results are checked to be independent of this choice. Robustness against
the location of the computational boundary is possible only if the mode is {\it trapped} (see below), 
which is a distinct feature of our calculations. By contrast,  if a density profile does not naturally trap modes (e.g., a power-law disk)
then the eccentricity eigenfunctions and eigenfrequencies will depend sensitively on the location of the computational boundary \citep{mir18}.}

In Figure~\ref{fig:eigenfunctions} (bottom panel), we show the fundamental mode eccentricity eigenfunctions $E$ for
 the values of $0.1\leq q_{\rm b}\leq1$ typically explored in CBD simulations.
These $E$ profiles peak within the cavity, and extend out to several times the binary separation $a_{\rm b}$.
However, at distances $R\gtrsim15 a_{\rm b}$, the eccentricity has decreased by 4 orders of magnitude, at which point the
disk can be considered to be effectively circular.

In Figure~\ref{fig:eigenfrequencies},
we show the eigenfrequencies $\omega_0$ of the fundamental mode as a function of $q_{\rm b}$ for
different values of $h_0$. These appear to roughly track 
the quadrupole precession frequency \citep[e.g.,][]{mor04} evaluated at $R_{\rm cav}$:
\begin{equation}\label{eq:characteristic_frequencies}
\omega_{\cal Q}\equiv \frac{3}{4}\frac{q_{\rm b}}{(1+q_{\rm b})^2}\left(\frac{R_{\rm cav}}{a_{\rm b}}\right)^{-2}\Omega_{\rm cav}~
\end{equation}
where $\Omega_{\rm cav}=\Omega\vert_{R=R_{\rm cav}}$ \citep{mac08}.
More  precisely, $\omega_0$  is suppressed relative  to 
$\omega_{\cal Q}$
by a modest
reduction factor that weakly depends on $h_0$.  Surprisingly, $\omega_0$ continues to track
$\omega_{\cal Q}$ at very low $q_b$, where one might have na\"ively expected pressure effects
to dominate.  We explore that behavior in more detail below.

Although there are  three adjustable parameters ($q_b$, $h_0$,  and $R_{\rm cav}$), 
the nature of the solutions is determined by  a single combination of them. Specifically, if we define
the pressure-induced precession frequency evaluated at $R_{\rm cav}$ via
\begin{equation}
\omega_{\cal P}\equiv h_0^2\Omega_{\rm cav}
\end{equation}
\citep{gol03,lee19a},
then
 the determining parameter is $\omega_{\cal P}/\omega_{\cal Q}$, the ratio of the pressure-induced
 precession rate to the quadrupole rate, evaluated at $R_{\rm cav}$. To show that, we
  solve the BVP for 1083 different sets of parameters, with $q_{\rm b}\in[0.003,1]$,
  $h_0\in\{0.01, 0.02, 0.05, 0.1, 0.2\}$ and $R_{\rm cav}\in\{2.0, 2.5, 3.0, 3.5,4.0, 4.5, 5.0\}$ ($l_0=0.7\Omega_{\rm b}a_{\rm b}^2$ is held fixed).
The resulting eigenfrequencies are depicted in Figure~\ref{fig:scaled_eigenfrequencies}, where
 we show the normalized eigenfrequency $\omega_0/\omega_{\cal Q}$ versus the ratio of characteristic frequencies $\omega_{\cal P}/\omega_{\cal Q}$ (color squares).
 As can be seen from the figure, the solutions nearly collapse into a single line\footnote{
 An exact collapse of the curves into a perfect line can be achieved by setting $l_0=0$ or by rescaling
 $l_0$ such that $l_0R_{\rm cav}^{-1/2}=$constant.}. 
 The ratio $\omega_{\cal P}/\omega_{\cal Q}$ also dictates the shape of the $E$-eigenfunctions.
 The eccentricity profiles associated to subset of the 1083 BVPs of Figure~\ref{fig:scaled_eigenfrequencies}
  is shown in Figure~\ref{fig:collapsed_eigenfunctions} (left panels).  When plotted
  as a function of $R/R_{\rm cav}$,  $E$-eigenfunctions with a given value of $\omega_{\cal P}/\omega_{\cal Q}$ line up with each other.
  When $\omega_{\cal P}/\omega_{\cal Q}\ll1$ (top), the
  eccentricity profile is confined to the immediate vicinity of the circumbinary cavity; when $\omega_{\cal P}/\omega_{\cal Q}\gg1$ (bottom)
  the eccentricity profile extends out to $R\gg R_{\rm cav}$.

\subsection{WKB Theory}\label{sec:wkb}

\subsubsection{Effective Potential}\label{sec:effective_potential}
WKB theory provides further physical insight into the numerical results presented above. We introduce a rescaled eccentricity, $y$, defined via
 $E=y(\Sigma R^3)^{-1/2}$,
which removes the first-order derivatives in the BVP \citep[e.g.][]{lanczos2012,gou07}, 
and results in
\begin{equation}\label{eq:eccentricity_scaled_form}
\frac{d^2 {y}}{d R^2}
+k^2y=0~,\;\;\; \text{with}\;\;
k^2(\omega,R)=\frac{2\Omega}{c_s^2}\left[\omega_{\rm pot}(R)-\omega\right]
\end{equation}
where  $\omega_{\rm pot}(R)$  is an ``effective potential'' in units of frequency \citep[e.g.,][]{lee19b}.
 Except for the additional factor $2\Omega/c_s^2$, Equation~(\ref{eq:eccentricity_scaled_form})
 is the time-independent Schr\"odinger equation in 1D \citep{ogi08}.
For an axisymmetric density profile $\Sigma(R)$, the effective potential is
\begin{equation}\label{eq:effective_potential}
\begin{split}
\omega_{\rm pot}(R)&=\Omega f_0+
\frac{h_0^2\Omega_{\rm b}}{2}
\left({\frac{R}{a_{\rm b}}}\right)^{\!\!\!-\tfrac{3}{2}}
\!\!\!\bigg[
\frac{R\Sigma'}{2\Sigma}+ \left(\!\frac{R\Sigma'}{2\Sigma}\!\right)^2 - \frac{R^2\Sigma''}{2\Sigma}-\frac{3}{4} \bigg]\\
&=\omega_{\cal Q}\left(\frac{R}{R_{\rm cav}}\right)^{-\frac{7}{2}}\\
&~~~~ +\frac{\omega_{\cal P}}{2}\left({\frac{R}{R_{\rm cav}}}\right)^{\!\!\!-\tfrac{3}{2}}
\!\!\!\bigg[\frac{R\Sigma'}{2\Sigma}+ \left(\!\frac{R\Sigma'}{2\Sigma}\!\right)^2 - \frac{R^2\Sigma''}{2\Sigma}-\frac{3}{4} \bigg]
\end{split}
\end{equation}
where primes denote radial derivatives.
The first term on the right hand side of Equation~(\ref{eq:effective_potential}) is the ``quadrupole contribution'' \citep[e.g.,][]{mor04}. 
The remaining terms are the ``pressure contribution,'' 
 which depend on the $\Sigma$ profile, but not its normalization. 
 
The sign of $\omega_{\rm pot}$ helps us determine
 the sign of $\omega$, i.e., whether modes are prograde or retrograde \citep[e.g.][]{tey16}.
For instance, the quadrupole contribution is the precession rate of a test particle around a binary and is always positive.
On the other hand, the pressure contribution is small and negative far from the cavity, as in for pure power-law disks \citep[e.g.,][]{gol03,mir18}, but
it becomes positive near the cavity edge.  As we  show below, all of our modes are prograde, even when pressure dominates over
the quadrupole.

\begin{figure*}
\centering
\includegraphics[width=0.95\textwidth]{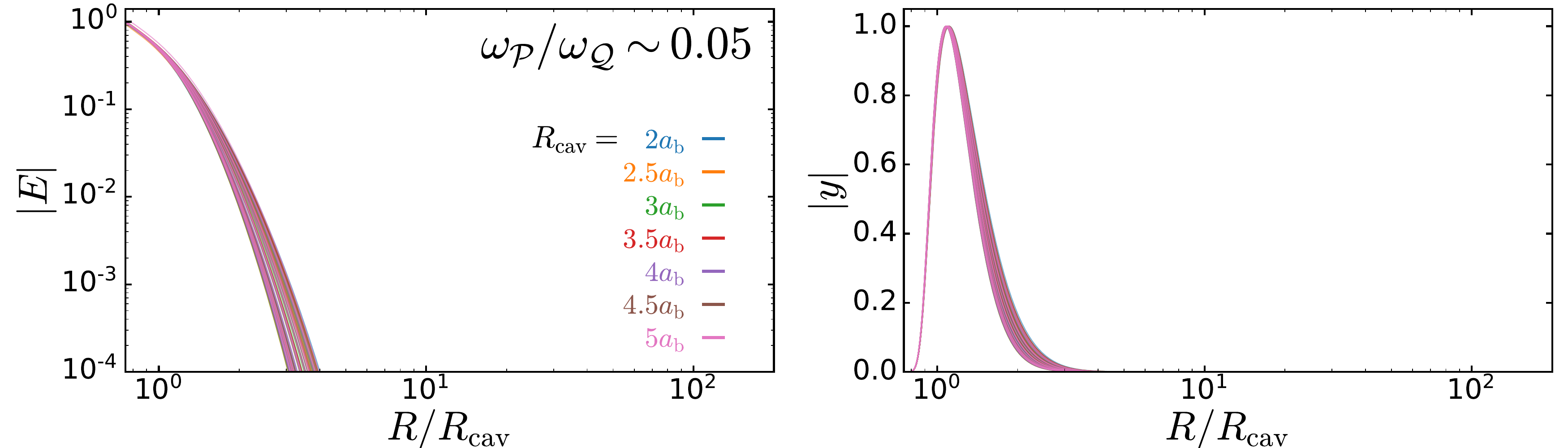}
\includegraphics[width=0.95\textwidth]{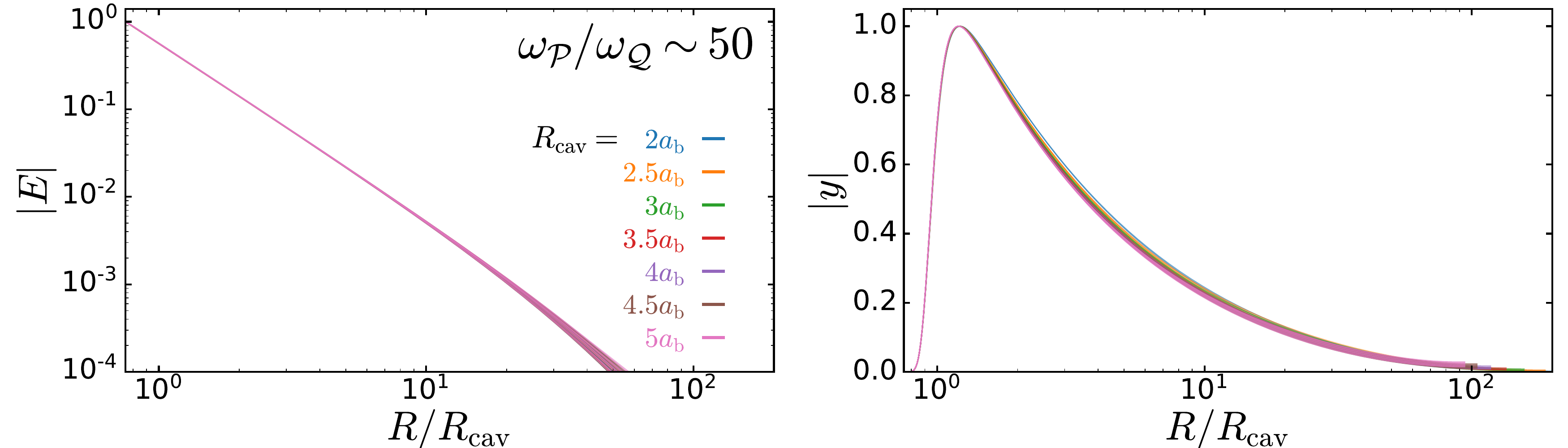}
\caption{Left panels: multiple eccentricity eigenfunctions $E$ obtained from the numerical solution to the BVP 
for $0.0375<\omega_{\cal P}/\omega_{\cal Q}<0.075$ (top) and  $37.5<\omega_{\cal P}/\omega_{\cal Q}<75$ (bottom).
Right panels: rescaled eccentricity  $|y|=|E|(\Sigma R^3)^{1/2}$ corresponding to the $E$-eigenfunctions in the right panels.
Color denotes the value of $R_{\rm cav}$ as in Figure~\ref{fig:scaled_eigenfrequencies}.
\label{fig:collapsed_eigenfunctions}}
\vspace{-0.0in}
\end{figure*}

When the circumbinary cavity
 is included, $\omega_{\rm pot}$ develops a local maximum, which is crucial for the existence of trapped modes.
 The function
 $\omega_{\rm pot}$ (depicted in Figure~\ref{fig:wave_diagram}, left panel) exhibits a general shape that
consists of a repulsive ($\omega_{\rm pot}\rightarrow-\infty$) inner region and attractive ($\omega_{\rm pot}>0$) ``potential well'' \footnote{
 Note that  
 it is the negative of $\omega_{\rm pot}$ that corresponds to the quantum mechanical potential.
 }, 
that peaks 
at $R_{\rm peak}= g R_{\rm cav}$ with $g$ of order unity.
The potential well can result in trapped modes (``bound states'') within that region provided that the peak is tall enough\footnote{
Asymmetric potential wells in 1D do not guarantee the existence of bound states \citep[][\S 22]{landau1981},
while symmetric wells always have a ground state \citep{sim76}.
}. 
The potential well is accompanied two turning points, beyond which waves cannot propagate. 
In the ``classically forbidden region'' depicted in gray in Figure~\ref{fig:wave_diagram},  the $y$-eigenfunctions are evanescent.
Oscillatory solutions are allowed outside the ``potential well'', but only in the form of traveling waves (``free states'') of negative frequency.

The trapping of the $y$-eigenfunctions can be seen in the right panels of Figure~\ref{fig:collapsed_eigenfunctions}. These profiles
are strongly-peaked functions with exponential cutoffs due to the left and right turning points (when $\omega=\omega_{\rm pot}$).
The left exponential cutoff is largely independent of $\omega_{\cal P}/\omega_{\cal Q}$; conversely, the right cutoff depends sensitively on
$\omega_{\cal P}/\omega_{\cal Q}$, with larger values producing more delocalized eigenfunctions.

\subsubsection{Approximate Eigenfrequencies}
The sharp local maximum in $\omega_{\rm pot}$ suggests that we can study eigenmodes trapped deeply
into the potential well by expanding to low order in $R$ and looking for known solutions of the time-independent
Schr\"odinger equation. First, we write Equation~(\ref{eq:eccentricity_scaled_form}) in ``Liouville
normal form''  \citep{liouville1837,amrein2005} to eliminate the pre-factor multiplying $\omega$.
With a change of variables $y=Y(R/R_{\rm cav})^{1/8}$, $\xi=(R/R_{\rm cav})^{3/4}$, we have
\begin{equation}\label{eq:eccentricity_liouville_form}
\frac{9\omega_{\cal P}}{32}\frac{d^2Y}{d\xi^2}+\left[\omega_{\rm pot}[R(\xi)]-\frac{7\omega_{\cal P}}{128}\xi^{-2} \right]Y=\omega Y
\end{equation}
Second, we expand the pressure-dependent term in $\omega_{\rm pot}$ 
around the local maximum $\xi_{\rm peak}$ to second order in $\xi$ and evaluate the remaining terms at $\xi=\xi_{\rm peak}$, i.e.,
the term in square bracket in Equation~(\ref{eq:eccentricity_liouville_form} is approximated by a quadratic potential.
Therefore, the resulting expression can be cast into the standard Schr\"odinger equation for the quantum harmonic oscillator (QHO), and thus
the eigenfrequencies in Equation~(\ref{eq:eccentricity_liouville_form}) are given by:
\begin{equation}\label{eq:eigenfrequency_approximate}
\omega_n^{\rm QHO}\approx 1.13\omega_{\cal Q}+26.1\omega_{\cal P}-130\omega_{\cal P}\left(n+\frac{1}{2}\right)~
\end{equation}
(see Appendix~\ref{app:b}). Equation~(\ref{eq:eigenfrequency_approximate}) is depicted in Figure~\ref{fig:scaled_eigenfrequencies} (thin black line)
for $n=0$. When $\omega_{\cal P}/\omega_{\cal Q}\ll1$, the analytic $\omega_0^{\rm QHO}$ and the BVP frequencies show a moderate level of agreement, 
indicating that the modes are rarely deep enough into the potential well to be properly described with this local expansion.

\subsubsection{WKB Eigenfrequencies from the Quantization Condition}\label{sec:quantum}
The {WKB method
 of elementary quantum mechanics can be
directly applied to obtain the solutions to Equation~(\ref{eq:eccentricity_scaled_form}), 
provided $k$ is sufficiently large compared to the lengthscale of variation of
$\omega_{\rm pot}$. Trapped modes are those with discrete eigenvalues $\omega_n$  that satisfy the
Einstein-Brillouin-Keller quantization condition \citep{ein17,kel58}, which is given by
\begin{equation}\label{eq:quantum_condition}
\oint k(\omega_n,R) dR =\left(2n+\frac{\mu}{2}\right)\pi
\end{equation}
where
$k$ is from the dispersion relation (Eq. \ref{eq:eccentricity_scaled_form}), and
 $\mu$ is the amount of phase loss, sometimes called the Maslov index\footnote{
The Maslov index corresponds to the number of turning points through a smooth potential (i.e., a soft reflection) 
plus  {\it twice} the number of turning points under Dirichlet boundary conditions (i.e. a hard reflection).}. 
In this case, $\mu=2$, which is the number of classical turning points for a 1D potential
like $\omega_{\rm pot}$  \citep{mar77,shu90,lee19a}.

\begin{figure*}
\centering
\includegraphics[width=0.98\textwidth]{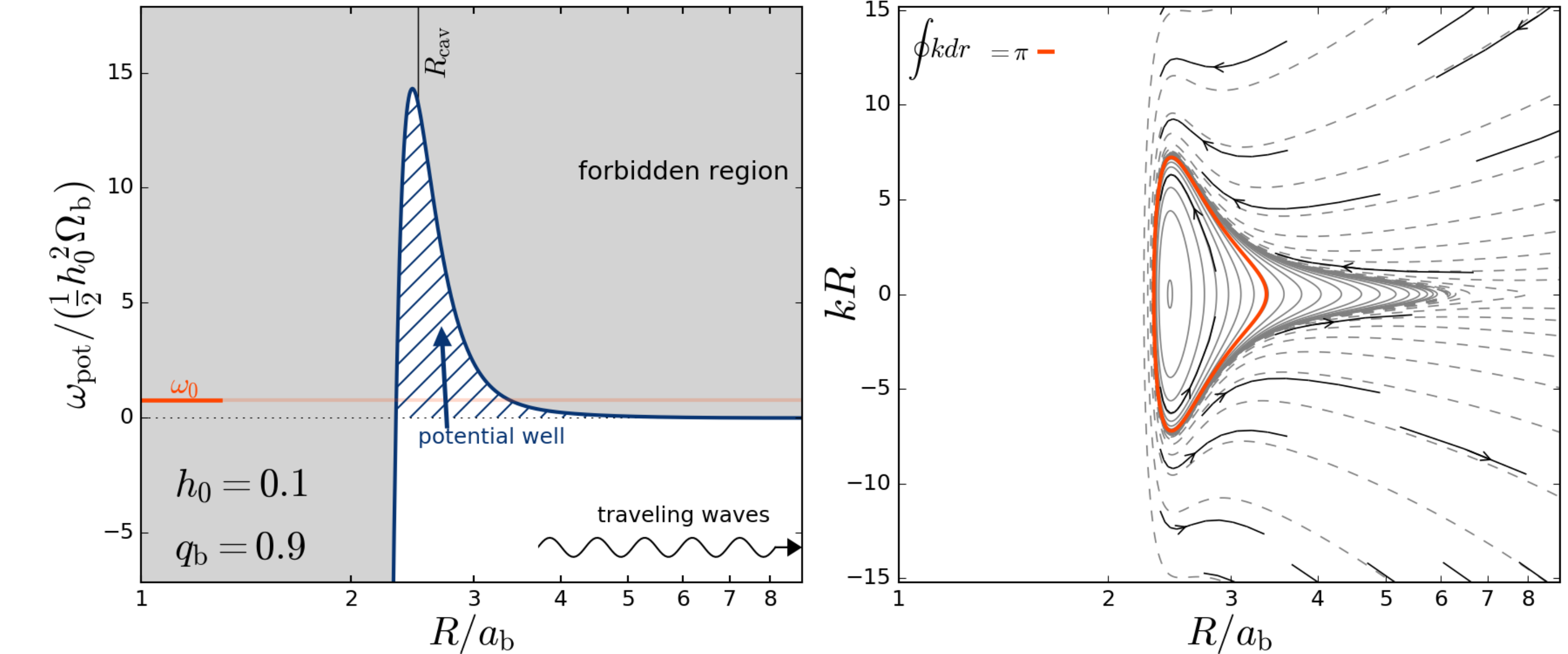}
\caption{Left panel: effective potential $\omega_{\rm pot}$ (blue curve, Equation~\ref{eq:effective_potential}) 
for fiducial parameters and
$h_0=0.1$ and $q_{\rm b}=0.9$ ($\omega_{\cal P}/\omega_{\cal Q}=0.33$).
Forbidden, trapped, and traveling-wave regions are labeled (see text).
The red line depicts the fundamental frequency $\omega_0$ that satisfies the 
quantization condition ~(\ref{eq:quantum_condition}) with $n=0$.
Right panel: WKB dispersion relation map (DRM) \citep{tre01,lee19a} depicting the the contours of constant frequency $\omega$ 
in $R-kR$ space.
Negative frequencies (dashed contours) allow for outgoing ($kR<0$) and incoming waves  ($kR>0$).
Positive frequencies (solid contours) exhibit two turning points, resulting in closed loops. The highlighted
contour in red satisfies $\oint kdR=\pi$.
\label{fig:wave_diagram}}
\vspace{-0.0in}
\end{figure*}

The behavior of a trapped mode is further illustrated in the dispersion relation map (DRM) of Figure~\ref{fig:wave_diagram} (right panel),
which depicts constant-$\omega$ contours
of the  dispersion relation (Equation~\ref{eq:eccentricity_scaled_form}) for different values of $R$ and $kR$. Traveling waves propagate
along open contours, while trapped modes (standing waves) trace closed loops. For the example of
the figure, the quantization condition is satisfied by one and only one such closed loop  (for $n=0$, in red).
The existence of just one trapped mode is a general property of
CBDs with $h_0\sim0.1$, which is the typical aspect ratio of 
of most CBD simulations \citep[e.g.,][]{far14,mir17,mun19,duf20}.
Colder disks ($h_0\lesssim0.05$), on the other hand,
can support higher-$n$ modes (Appendix~\ref{app:a}).  This type of disk
has been simulated recently \citep{thu17,tie20}, but no multi-harmonic
disk eccentricity has been reported.
 
 Figure~\ref{fig:scaled_eigenfrequencies} also includes the WKB frequencies as a function of  $\omega_{\cal P}/\omega_{\cal Q}$.
 The agreement with the BVP frequencies is excellent for $\omega_{\cal P}/\omega_{\cal Q}<1$, but the two solution quickly
 diverge when  $\omega_{\cal P}/\omega_{\cal Q}>1$, i.e., at low eigenfrequencies. To understand this discrepancy, we turn to studying the WKB eigenfunctions
 below.

 \subsubsection{Approximate Eccentricity Profiles}
Far from the cavity, we can write the evanescent part of the $y$-eigenfunction using the WKB approximation
$y{\sim}\exp[-\int dR (-k^2)^{1/2}]$. For $\omega\gg|\omega_{\rm pot}|$, we obtain
\begin{equation}\label{eq:approximate_eigenfunction}
E\sim R^{-5/4} \exp\left[-\left(\frac{R}{\lambda}\right)^{3/4}\right]
\end{equation}
where
\begin{equation}\label{eq:tapering_radius}
\lambda
=R_{\rm cav}\left(\frac{9}{32}\frac{\omega_{\cal P}}{\omega_0}\right)^{2/3}
 \end{equation}
i.e., the eccentricity profile is a tapered power-law with tapering radius $\lambda$ \citep[see also appendix C in][]{shi12}.
{A rough approximation of the quantization condition
gives $\lambda\sim R_{\rm cav}({\omega_{\cal P}}/{\omega_{\cal Q}})^{2/3}[1-0.02({\omega_{\cal Q}}/{\omega_{\cal P}})]$
(Appendix~\ref{app:c}).}
Equation~(\ref{eq:approximate_eigenfunction}) is included in Figure~\ref{fig:eigenfunctions} (for clarity, only for the $q_{\rm b}=0.1$ case).
The tapering effect
is a natural consequence of the $y$-eigenfunction  being a trapped mode.
Note that, in Equation~(\ref{eq:tapering_radius}), $\lambda$ is arbitrarily large for arbitrarily small $\omega$,
which is the usual behavior of marginally bound states in quantum mechanical potential wells.
This ``delocalization'' of the eigenfunctions explains the radial extent of $E$ and $y$
 in Figure~\ref{fig:collapsed_eigenfunctions}, in turn explaining why WKB fails at low frequencies, since
 the spatial wave number is too small for the approximation to hold. As a consequence, when $\omega_{\cal P}/\omega_{\cal Q}\rightarrow\infty$
 (i.e., when the quadrupole contribution is negligible), eigenfunctions are not 
trapped, even if the CBD is steeply truncated around the binary.
 
We may now also qualitatively understand the  
 surprising result that $\omega_0\sim \omega_{\cal Q}$ (Figures \ref{fig:eigenfrequencies} and \ref{fig:scaled_eigenfrequencies}), even  when 
 when pressure naively dominates over the quadrupole ($\omega_{\cal P}/ \omega_{\cal Q}\gg 1$).
 The  reason behind this result is that, when $\omega_{\cal P}/\omega_{\cal Q}$ is large, the delocalized eigenfunction extends out to many 
 times $R_{\rm cav}$,  in which case the pressure-induced precession rate is much lower than the naive expectation $\omega_{\cal P}\sim h_0^2\Omega_{\rm cav}$. 
 Instead, the magnitude of the pressure-induced precession is set by the value of $h_0^2\Omega$ at $R\gg R_{\rm cav}$, where
 the eigenfunction is within a factor of a few of its peak value. At these distances, $h_0^2\Omega$ is much less 
 than $\omega_{\cal P}$, and can never overcome quadrupole-induced
 precession.


\begin{figure}[b!]
\includegraphics[width=0.44\textwidth]{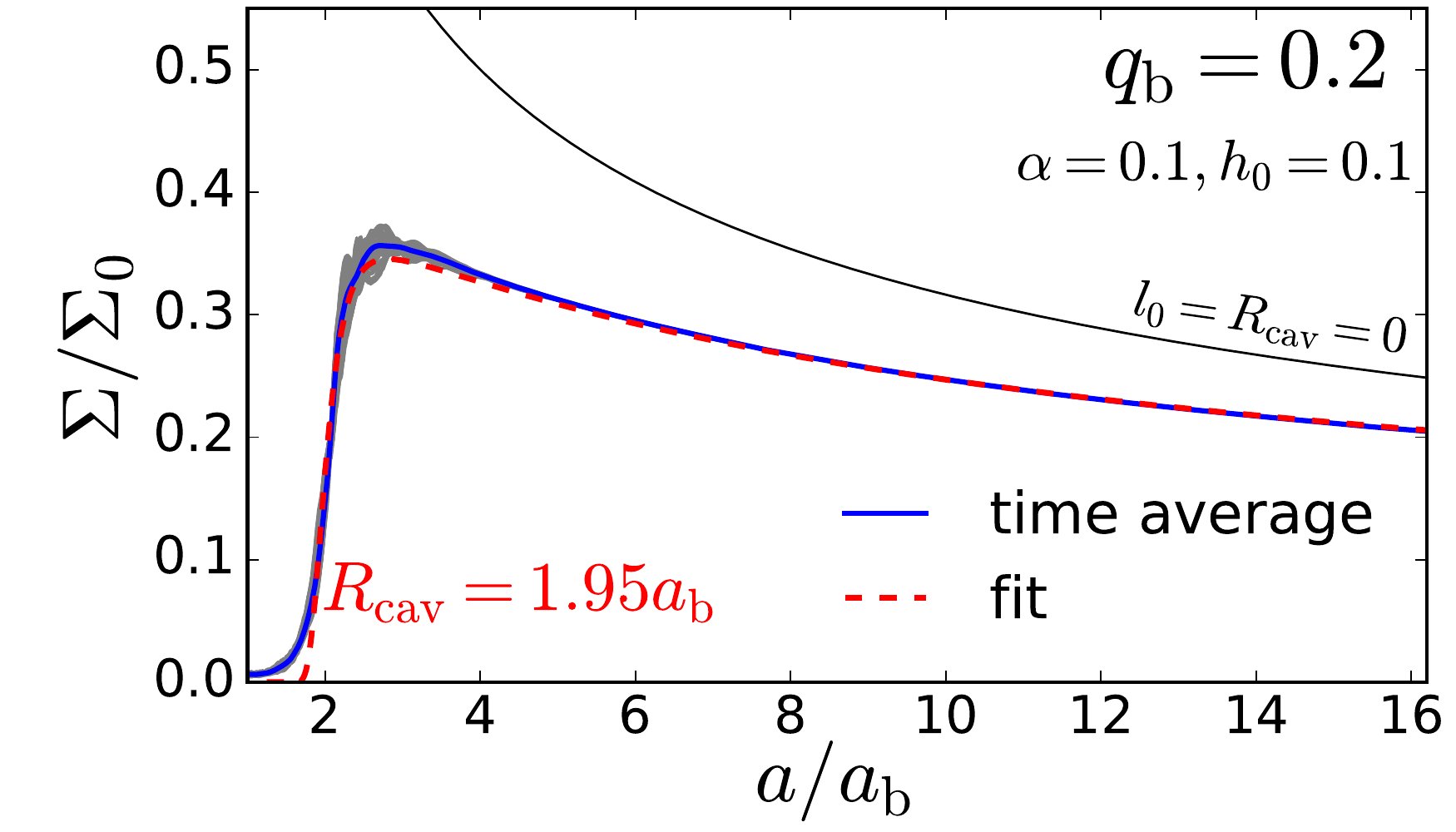}
\includegraphics[width=0.44\textwidth]{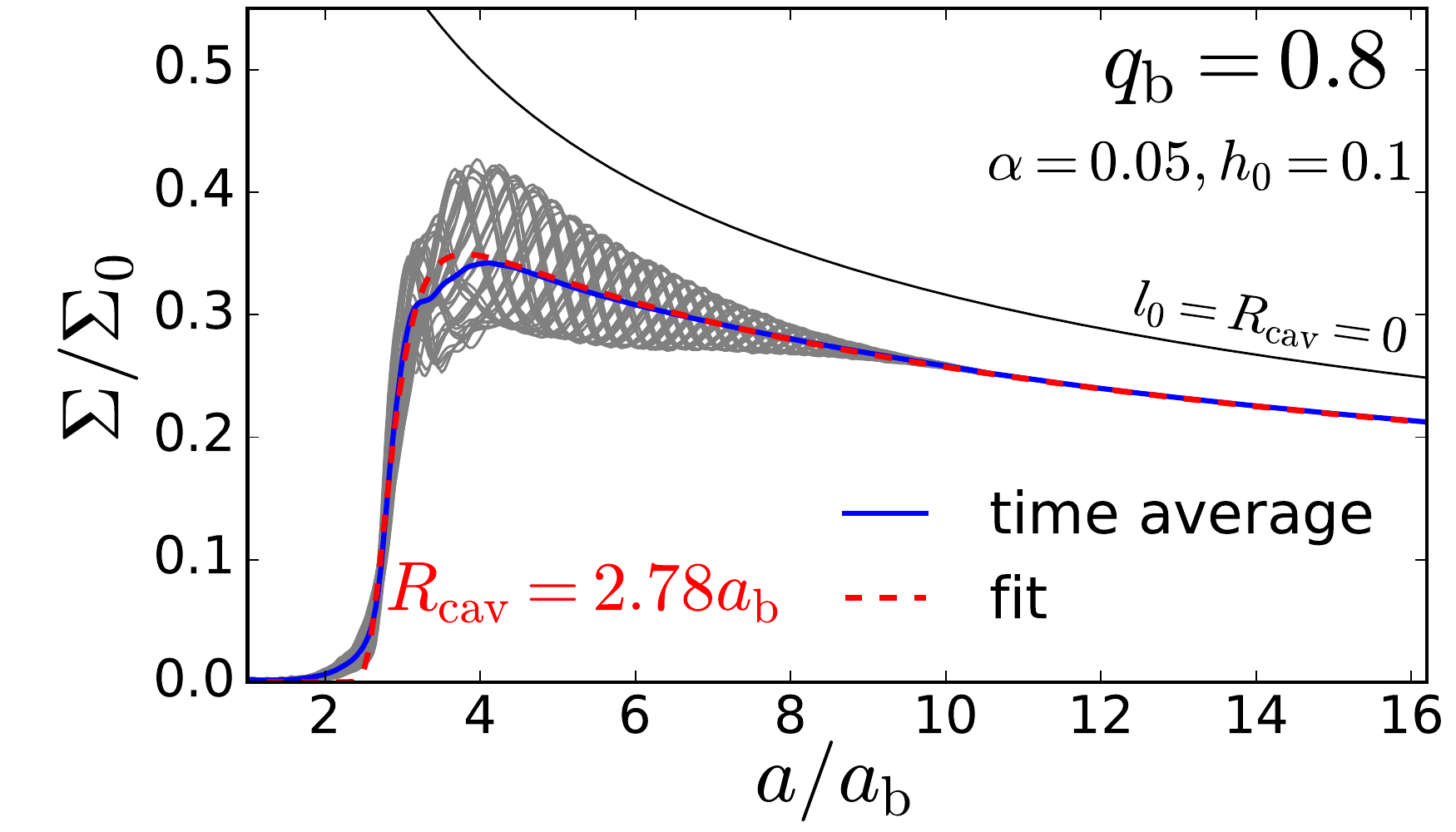}
\caption{Examples of  the time-averaged surface density profiles of CBDs in VSS, where the radial coordinate
is the semimajor axis ratio $a/a_{\rm b}$. Top panel: $\Sigma$ profile of a disk  with $\alpha=h_0=0.1$ around
a circular binary of mass ratio $q_{\rm b}=0.2$. Gray curves depict individual
snapshots of the density field; their time average (over 500 binary orbits) is depicted by the blue curve;
the dashed red line is the best-fit model (Equation~\ref{eq:density_profile}). Bottom panel: same as above
but for $q_{\rm b}=0.8$. In both panels, the thin black line depicts the power-law profile
$\Sigma=\Sigma_0 (a/a_{\rm b})^{-1/2}$, i.e., the torque-free solution with 
$l_0=R_{\rm cav}=0$ at the same accretion rate $\dot{M}_0$. As is well-known, positive values of $l_0$ are
tantamount to a density deficit relative to the power-law solution.
\label{fig:surface_density}}
\end{figure}

\section{Comparison to Hydrodynamical Simulations}\label{sec:nonlinear}

\subsection{Hydrodynamics of Circumbinary Accretion}\label{sec:hydro_disks}
We carry out 2D hydrodynamics simulations of steady-state CBDs 
using the moving-mesh code {\footnotesize AREPO} \citep{spr10a,pak16}
in its Navier-Stokes formulation \citep{mun13a}, using a locally isothermal
equation of state ($c_s^2\propto R^{-1}$) and an $\alpha$-viscosity prescription.
Once transients die out, these simulations are fully determined
by four parameters: $q_{\rm b}$, $e_{\rm b}$, $h_0$ and the viscosity coefficient $\alpha$.
We focus on the case $e_{\rm b}=0$
and vary the other parameters, restricting ourselves to the regime with $h_0\sim0.1$,
which, from WKB analysis, is expected to develop a single trapped mode.
We note that $\alpha$ does not appear in the linear calculations (Section~\ref{sec:linear}).

\begin{figure*}
\includegraphics[width=0.96\textwidth]{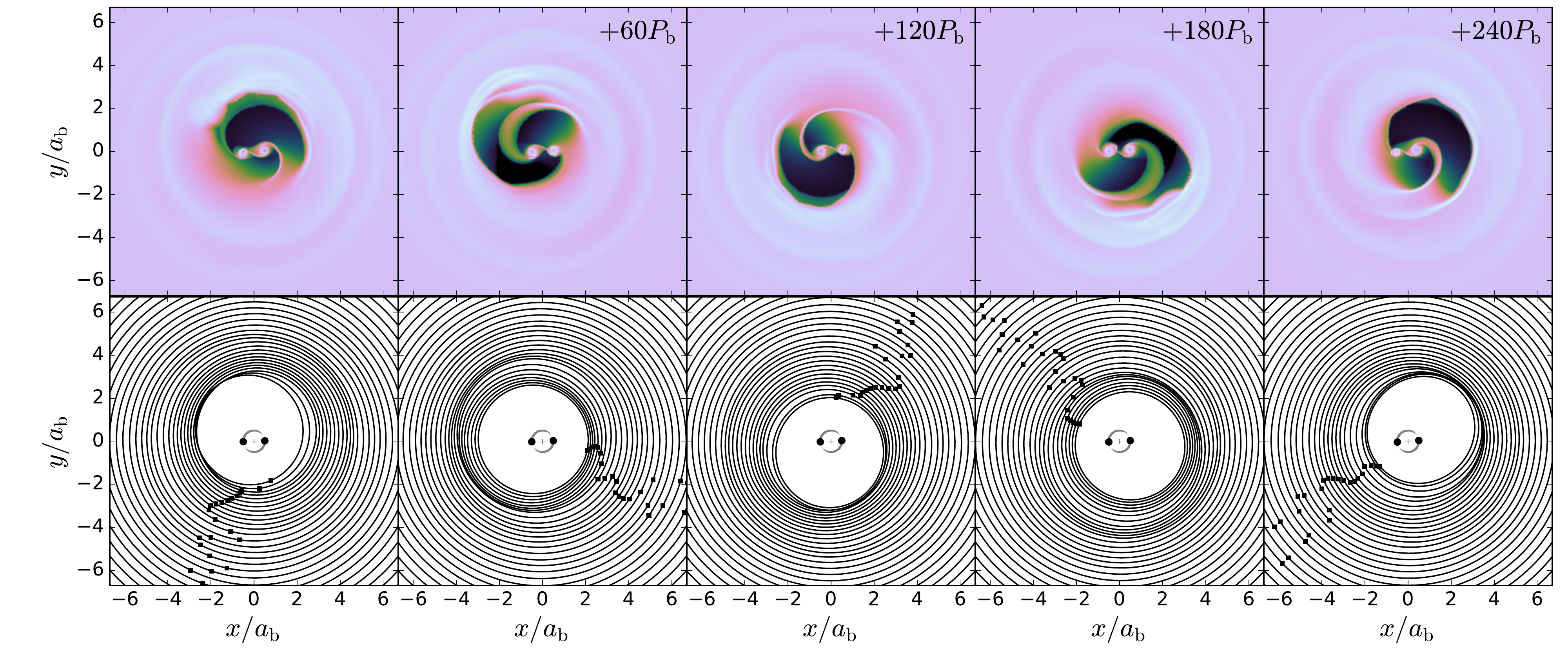}
\caption{Evolution of CBDs over secular timescales. Top panels: surface density (logarithmic scale) in the vicinity
of a binary with $q_{\rm b}=1$, $e_{\rm b}=0$ (disk parameters are $h_0=\alpha=0.1$) in intervals of 60 binary orbits
once steady-state has been achieved. The gas morphology is consistent throughout the panels, except for the orientation
of the central cavity, which evolves in tandem with the disk eccentricity.
Bottom panels: barycentric elliptical ``orbits'' corresponding to gas eccentricity (Equation~\ref{eq:eccentricity_vector})  binned in semi-major axis,
These ``orbits'' change in time, exhibiting prograde apsidal precession, as evidenced by the advancement of the longitude of pericenter $\varomega_{\rm d}$,
 depicted by solid black squares; the orientation of the ellipses is roughly coherent ($\varomega_{\rm d}$ is approximately equal for all radii) out to
a distance of $\sim 10 a_{\rm b}$ from the barycenter.  For this simulation, a fit of the parametric model  (\ref{eq:density_profile}) to the surface density 
gives $R_{\rm cav}=2.42a_{\rm b}$.
\label{fig:elliptical_orbits}}
\end{figure*}

\begin{figure*}
\centering
\includegraphics[width=0.96\textwidth]{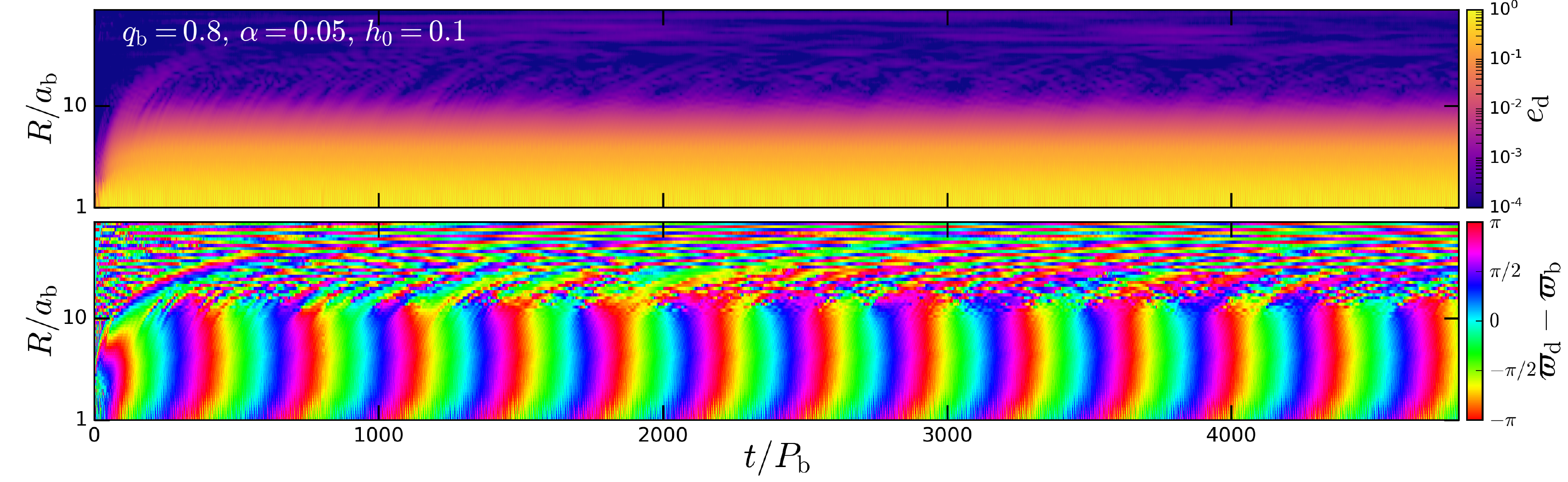}
\caption{Eccentricity evolution maps for a CBD with $h_0=0.1$ and $\alpha=0.05$ around a circular binary 
with $q_{\rm b}=0.8$.  For this simulation, a fit of the parametric model  (\ref{eq:density_profile}) to the surface density 
gives $R_{\rm cav}=2.78a_{\rm b}$.
\label{fig:eccentricity_diagram}}
\end{figure*}

Details on the model setup and numerical
scheme can be found in \citet{mun16b}, \citet{mun19} and \citet{mun20a}. The general
findings of the aforementioned works include:
\begin{itemize}
\item A boundary condition with constant mass supply  $\dot{M}_0$ enables CBDs to reach
viscous steady state  (VSS)  (see also \citealp{mir17,dem20a}).
Once VSS is reached, the binary accretion rate $\dot{M}_{\rm b}$ equals (on average) the supply rate $\dot{M}_0$.

\item In VSS, the net angular momentum current in the CBD $\dot{J}_{\rm d}$ (including advective, viscous and gravitational contributions) is statistically 
stationary and independent of radius \citep{mir17}. In addition, the time average $\langle\dot{J}_{\rm d}\rangle$ equals the net angular momentum transfer rate
from the disk onto the binary $\langle \dot{J}_{\rm b}\rangle$ (which is comprised of gravitational and accretional torques).

\item  Most importantly, the quantity $l_0\equiv \langle \dot{J}_{\rm b}\rangle/ \langle \dot{M}_{\rm b}\rangle$ is a positive constant. 
If $l_0>l_{0,{\rm crit}}$
\citep[where $l_{0,{\rm crit}}>0$ depends on the binary properties;][]{mun20a}, then the accretion process leads to binary expansion.
\end{itemize}

CBD simulations in VSS and positive $l_0$ produce stationary $\Sigma$ profiles that exhibit mass {\it deficits} in relation
to the power-law profiles of equal accretion rate\footnote{
This deficit is also a well-known consequence of the (misleadingly labeled) ``zero-torque
boundary  condition'' at some finite radius
\citep{lyn74,frank2002,dem20b}. Recently, however, \citet{tie20} has reported {\it pileups}, rather than deficits,
in simulations of cold ($h_0\lesssim0.04$) CBDs, which translate into negative values of $l_0$.
}. 
 Planetary-mass companions, by contrast, have $l_0<0$ and exhibit mass pile-ups \citep[e.g.,][]{dem20a}.
These stationary $\Sigma$ profiles are well described by Equation~(\ref{eq:density_profile}), which is a function
of three parameters: $\Sigma_0$, $l_0$ and $R_{\rm cav}$ ($\zeta=12$ is held fixed). 
Moreover, in VSS, 
$\Sigma_0 = {\dott{M}_0}/({3\pi \alpha h_0^2\Omega_{\rm b}a_{\rm b}^2})$, and $l_0$ can be computed
from the torque balance in the simulations; hence, the only
remaining parameter  for the $\Sigma$ profile is the cavity size
$R_{\rm cav}$, which we fit from time-averaged density profiles.

In Figure~\ref{fig:surface_density}, we show examples of the measured $\Sigma$ (in blue)
contrasted to the parametric model (Equation~\ref{eq:density_profile}, in red) with best-fit values of $R_{\rm cav}$
for two different simulations with $h_0=0.1$ ($q_{\rm b}=0.2$, $\alpha=0.1$ on top, and $q_{\rm b}=0.8$, $\alpha=0.05$ at the bottom).
With a fully determined $\Sigma$ profile and sound-speed profile, we can compute the linear eigenfunctions and
eigenfrequencies for a suite of hydrodynamical models (see Section~\ref{sec:comparison} below).

\subsection{Freely Precessing Eccentric Disks}\label{sec:simulation_results}
CBD simulations are known to develop lopsided cavities which change orientation on time-scales much longer
than the binary's orbital period \citep[e.g.,][]{mac08,mir17,thu17}. This slowly varying lopsidedness can be readily appreciated in 
 in Figure~\ref{fig:elliptical_orbits} (top panels), which shows the surface density every 60 binary orbits
for total time interval of $240$ binary orbits.

One can compute a ``dynamic'' eccentricity from the velocity field \citep[e.g.,][]{mac08}.
As in \citet{mir17}, we measure disk eccentricity by first computing the Laplace-Runge-Lenz
vector of the $i$-th gas cell  \citep[see also][]{tey17}
\begin{equation}\label{eq:eccentricity_vector}
{\bf e}_{i}=\frac{1}{\Omega_{\rm b}^2a_{\rm b}^3}{\bf v}_i\times({\bf r}_i\times{\bf v}_i)
-\frac{{\bf r}_i}{|{\bf r}_i|}~~.
\end{equation}
which can be combined into a global eccentricity vector ${\bf e}_{\rm d}$.
By binning cells in semi-major axis, we can construct a set of co-focal Keplerian orbits, as shown
in the bottom panels of Figure~\ref{fig:elliptical_orbits}. These ellipses match the orientation and precession 
rate of the lopsided cavity. More importantly, the ellipses exhibit apsidal coherence (the longitudes
of pericenter are aligned), meaning that they
precess in tandem, which is suggestive of proper mode behavior.

The time evolution of the disk eccentricity can be better assessed from space-time eccentricity maps.
Following \citet{mir17}, we bin the gas eccentricity vector ${\bf e}_i$ in barycentric radius to obtain ${\bf e}_{\rm d}$
as a function of $R$ and $t$. We visualize the
eccentricity $e_{\rm d}=(e_{{\rm d},x}^2+e_{{\rm d},y}^2)^{1/2}$ (top) and the longitude of pericenter $\varomega_{\rm d}=\tan^{-1}(e_{{\rm d},x},e_{{\rm d},y})$ (bottom)
as intensity maps. The maps of Figure~\ref{fig:eccentricity_diagram} correspond to a simulation with $q_{\rm b}=0.8$, $h_0=0.1$
and $\alpha=0.05$ integrated for $\simeq5000$ binary orbits.  Eccentricity growth is rapid: the 
 $e_{\rm d}$ map saturates after $\simeq 200P_{\rm b}$ and remains time-independent afterward. 
The  $\varomega_{\rm d}$ map achieves
a regularly repeating pattern after a few hundred binary orbits, showing that the disk precesses continuously, spanning the range
[0,$2\pi$] at a fixed rate. 

One can further define a global longitude of pericenter $\widebar{\varomega}_{\rm d}\equiv \tan^{-1} (\widebar{e}_{{\rm d},x},\widebar{e}_{{\rm d},y} )$
(Fig.~\ref{fig:eccentricity_diagram}, bottom panel)
where $\widebar{e}_{x,y}\equiv {\int \Sigma e_{x,y}  R dR}/({\int \Sigma R dR})$ denotes a mass-weighted radial average of the eccentricity
vector over the entire disk. The global precession frequency,
\begin{equation}\label{eq:disk_precession}
\dot{\varomega}_{\rm d}\equiv \frac{d}{dt} \widebar{\varomega}_{\rm d} ~,
\end{equation}
in the example of the figure, is $\dot{\varomega}_{\rm d}=2.77\times10^{-3}\Omega_{\rm b}$. Hydrodynamical simulation
typically exhibit precession rates $\sim{\cal O}(10^{-3}\Omega_{\rm b})$ \citep{mac08,mir17}, which is consistent
with the quadrupole-induced precession rate at the edge of the cavity.

\subsection{Comparison of  Theory and Simulations}\label{sec:comparison}
As a test of the validity of the linear eccentricity equation (\ref{eq:eccentricity_equation}),
we compare the hydrodynamic eccentricity profiles and precession rates to the
eigenfunctions and eigenfrequencies, respectively, obtained from solving the BVP.
The WKB
approximation is not expected to hold, since 
our simulations have $\omega_{\cal P}/\omega_{\cal Q}\simeq0.3-0.4$, which is slightly outside the regime
in which the WKB approximation is valid (see Figure~\ref{fig:scaled_eigenfrequencies}). 

Figure~\ref{fig:comparison_eigenfunction} shows
the eccentricity profiles obtained from hydrodynamics (squares) and the linear BVP (lines)
for $q_{\rm b}=$0.2, 0.4 and 1 and three different combinations
of $h_0$ and $\alpha$.  In each case, the values $R_{\rm cav}$ and $l_0$ used in the BVP are extracted from simulations.
The profiles show good agreement, revealing that the tapered power-law
behavior of $|E|$ (Equation~\ref{eq:approximate_eigenfunction}) is representative of
the eccentricity behavior in hydrodynamical simulations. The difference between the panels is subtle, although the dependence on viscosity is clear: 
in general,  a higher $\alpha$ corresponds to a smaller $R_{\rm cav}$ \citep[e.g.,][]{art94,mir15}.
Consequently, high viscosity shifts the eccentricity profile inward (see Equation~\ref{eq:approximate_eigenfunction}). 

Similarly, we can compare the empirical precession rate $\dot{\varomega}_{\rm d}$ (Equation~\ref{eq:disk_precession})
to the linear eigenfrequency $\omega_0$. In general,
the linear frequencies are within around 50\% of
 the hydrodynamical ones, and the high viscosity case (middle panel) shows
a remarkable agreement between the two approaches.

\begin{figure}
\centering
\includegraphics[width=0.4\textwidth]{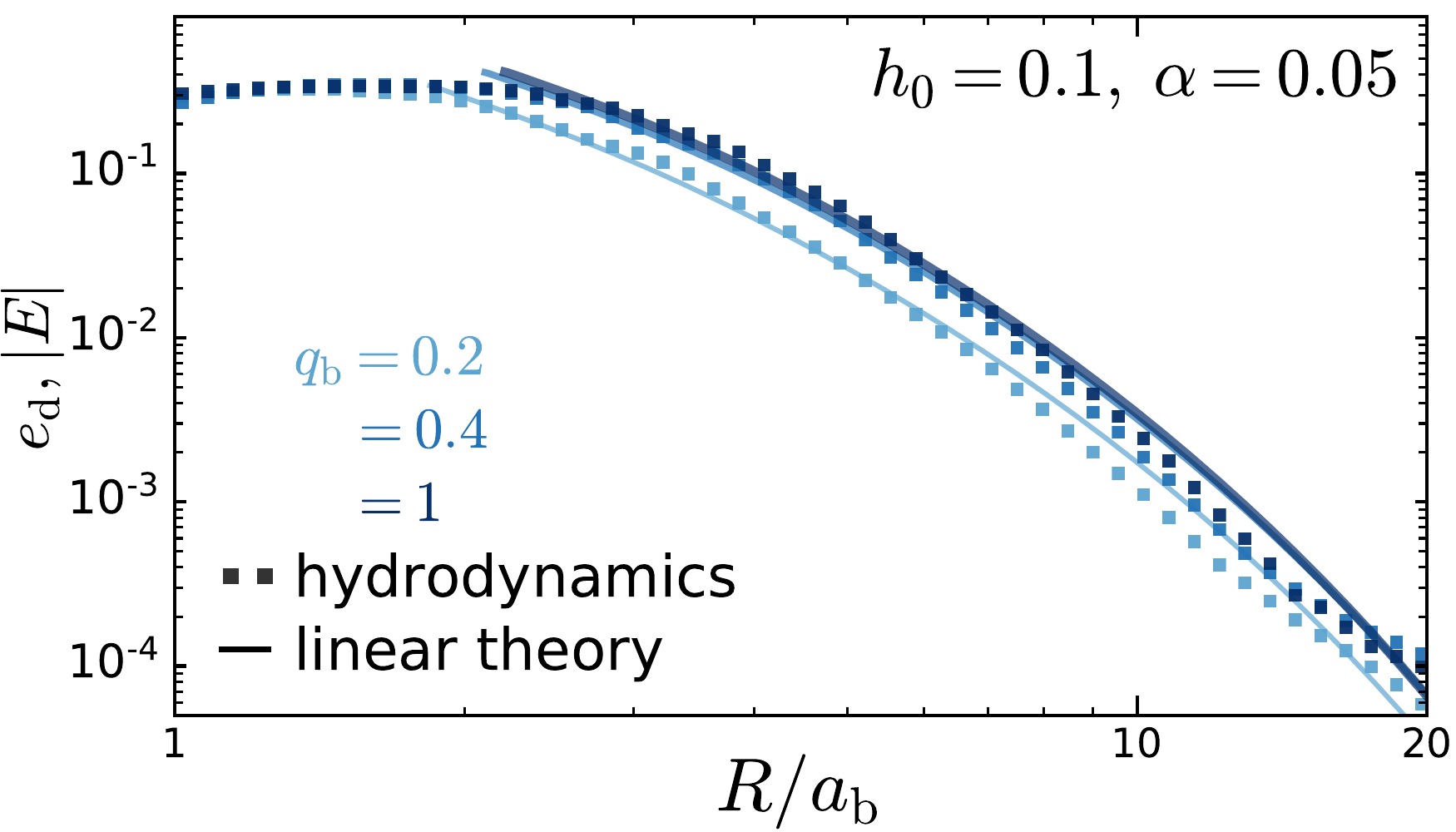}
\includegraphics[width=0.4\textwidth]{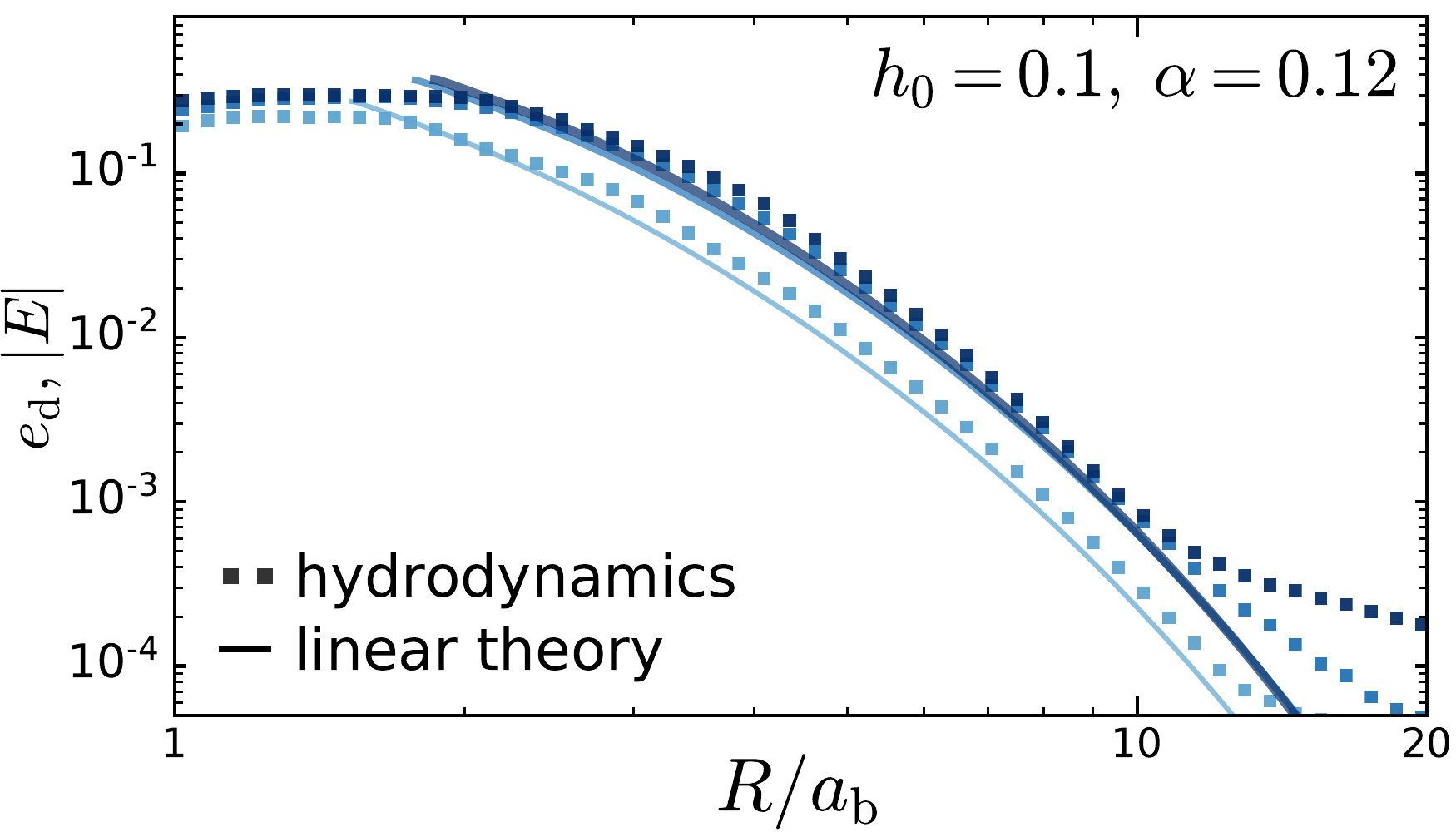}
\includegraphics[width=0.4\textwidth]{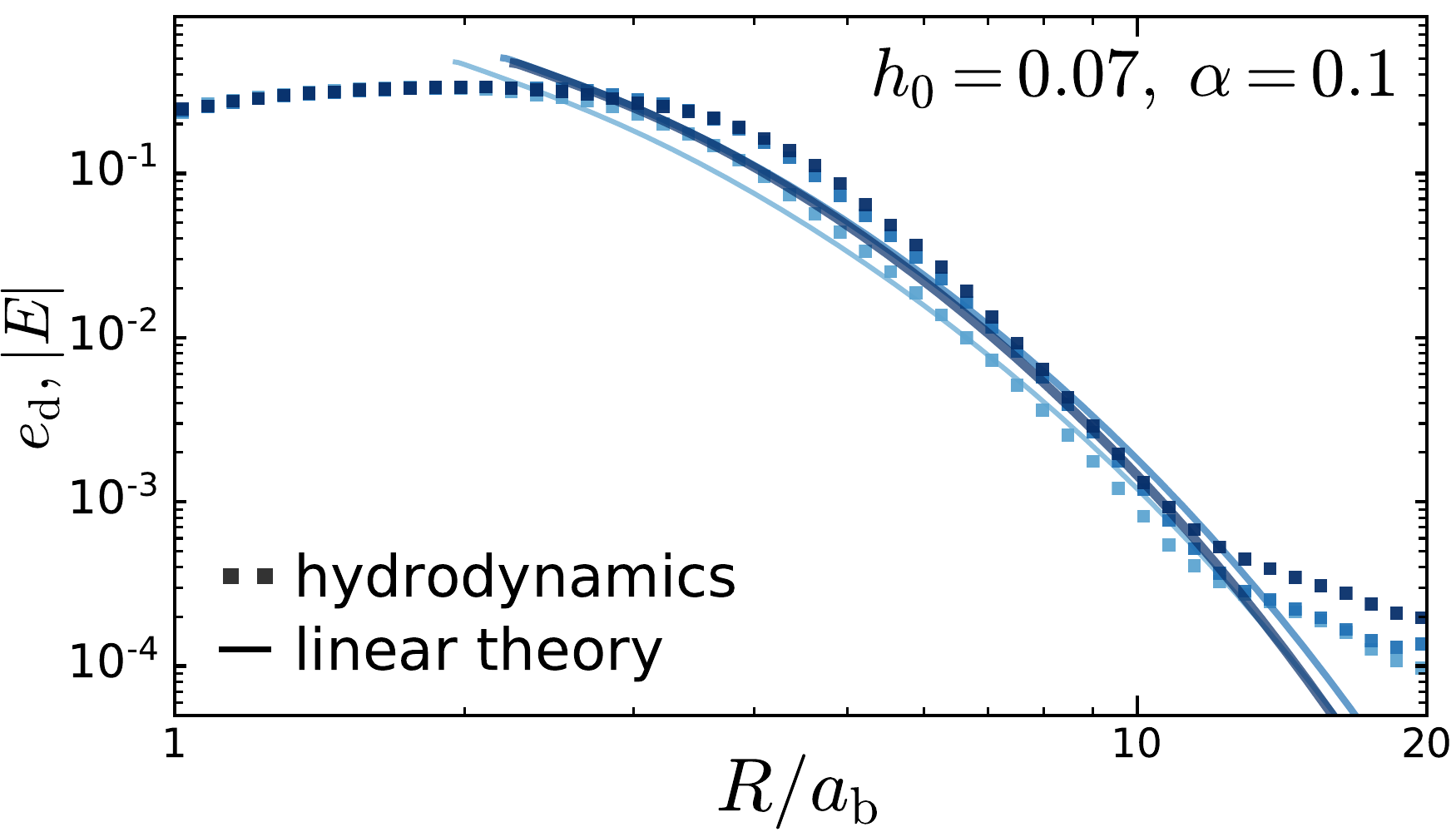}
\caption{ Eccentricity eigenfunctions $E$ derived from linear analysis (solid lines) compared to the empirical eccentricity
profiles $e_{\rm d}$ measured directly from hydrodynamical simulations (small square markers) as described in Section~\ref{sec:simulation_results}.
Each panel consists of three different binary mass ratios ($q_{\rm b}=0.2$, 0.4 and 1, from dark blue to light blue)
and panels differ in their combination of $h_0$ and $\alpha$. 
For each set of mass ratios $q_{\rm b}=(0.2,0.4,1)$ the cavity sizes are 
$R_{\rm cav}=(2.32a_{\rm b},~2.64a_{\rm b},~2.79a_{\rm b})$ (top),
$R_{\rm cav}=(1.90a_{\rm b},~2.22a_{\rm b},~2.33a_{\rm b})$ (middle),
 and $R_{\rm cav}=(2.48a_{\rm b},~2.78a_{\rm b},~2.86a_{\rm b})$ (bottom).
 The normalization of the linear eigenfunctions is arbitrary, and is chosen to match
the hydrodynamical eccentricity at the edge of the cavity.
\label{fig:comparison_eigenfunction}}
\end{figure}

\begin{figure}
\includegraphics[width=0.4\textwidth]{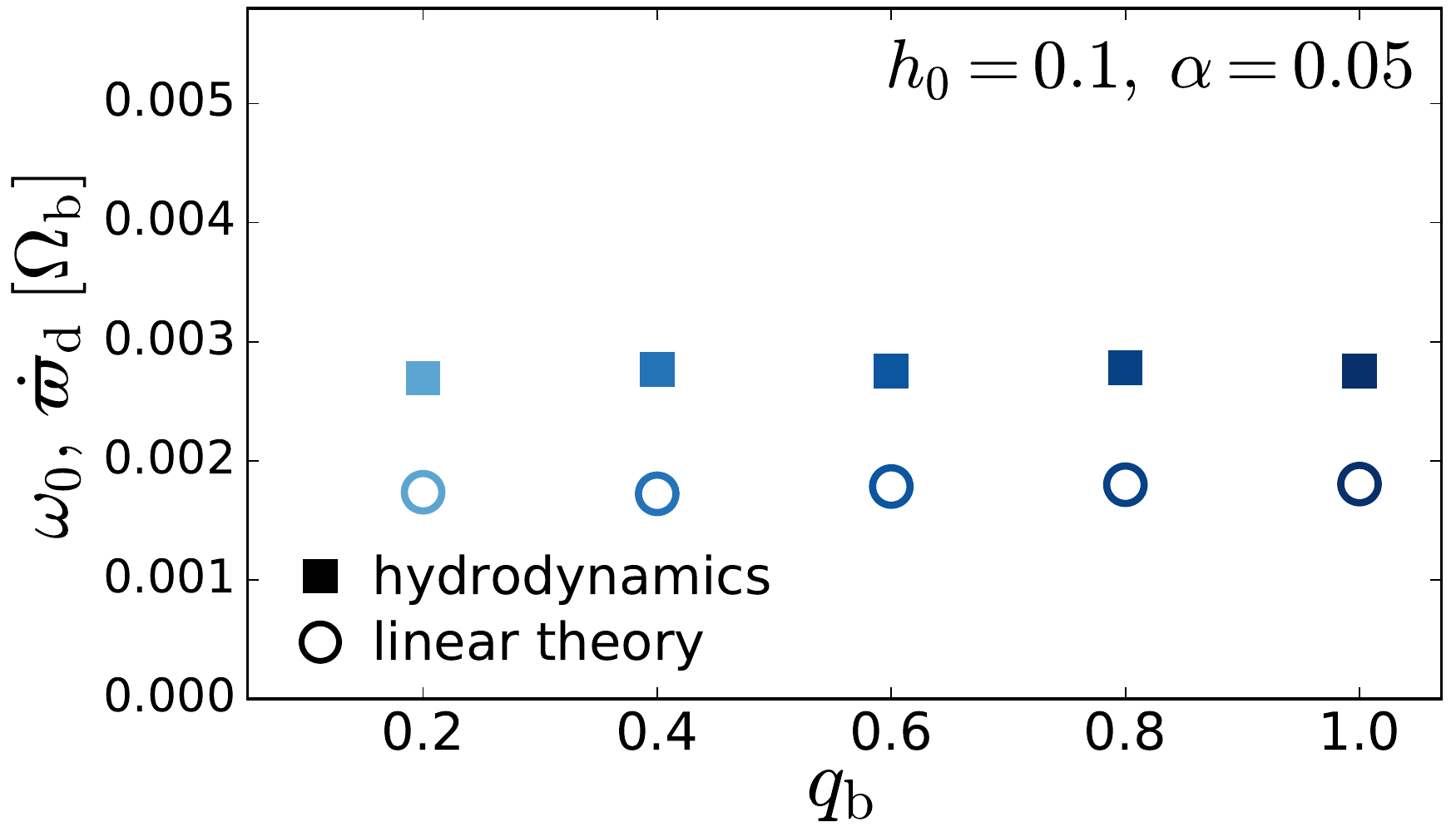}
\includegraphics[width=0.4\textwidth]{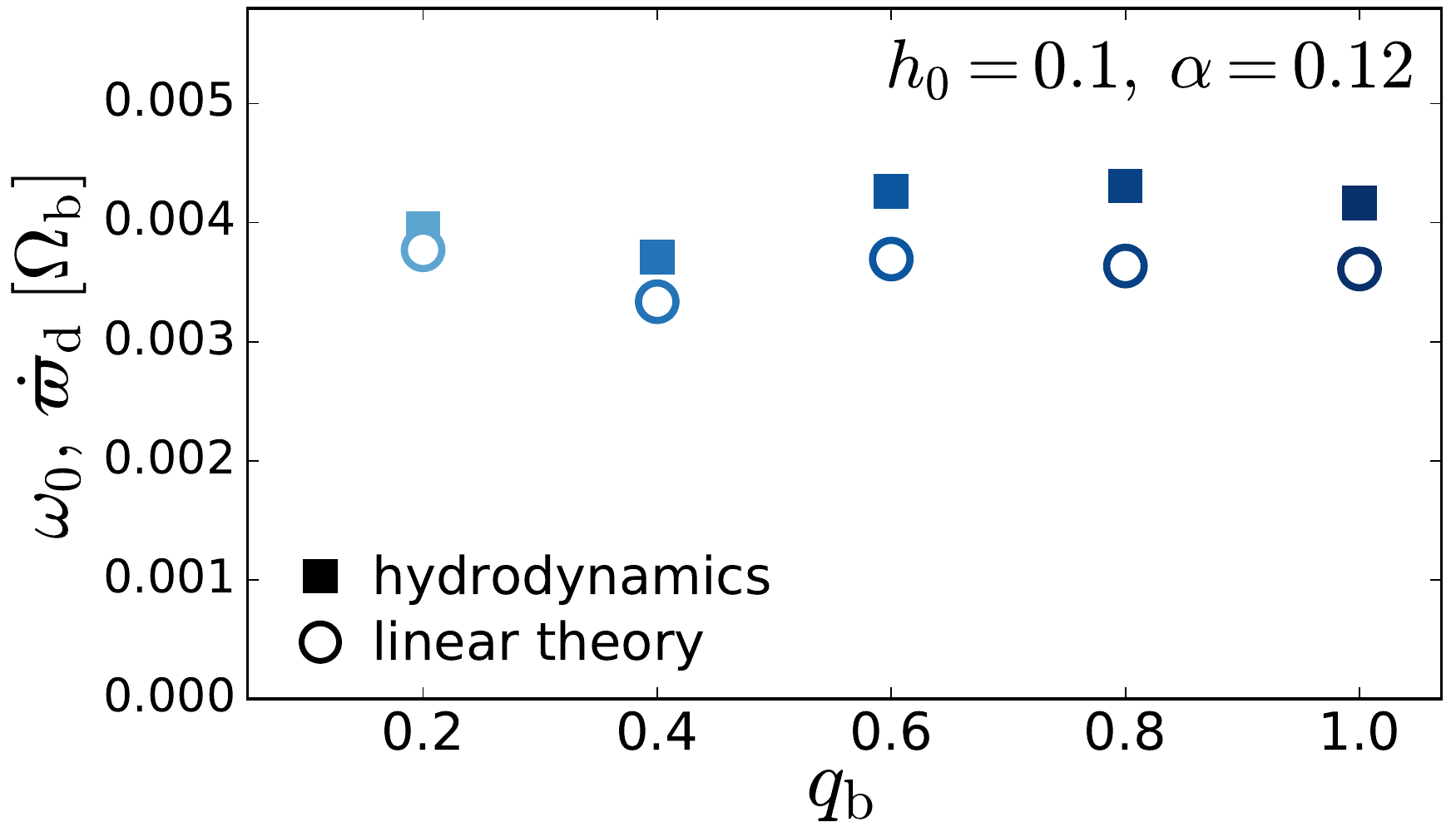}
\includegraphics[width=0.4\textwidth]{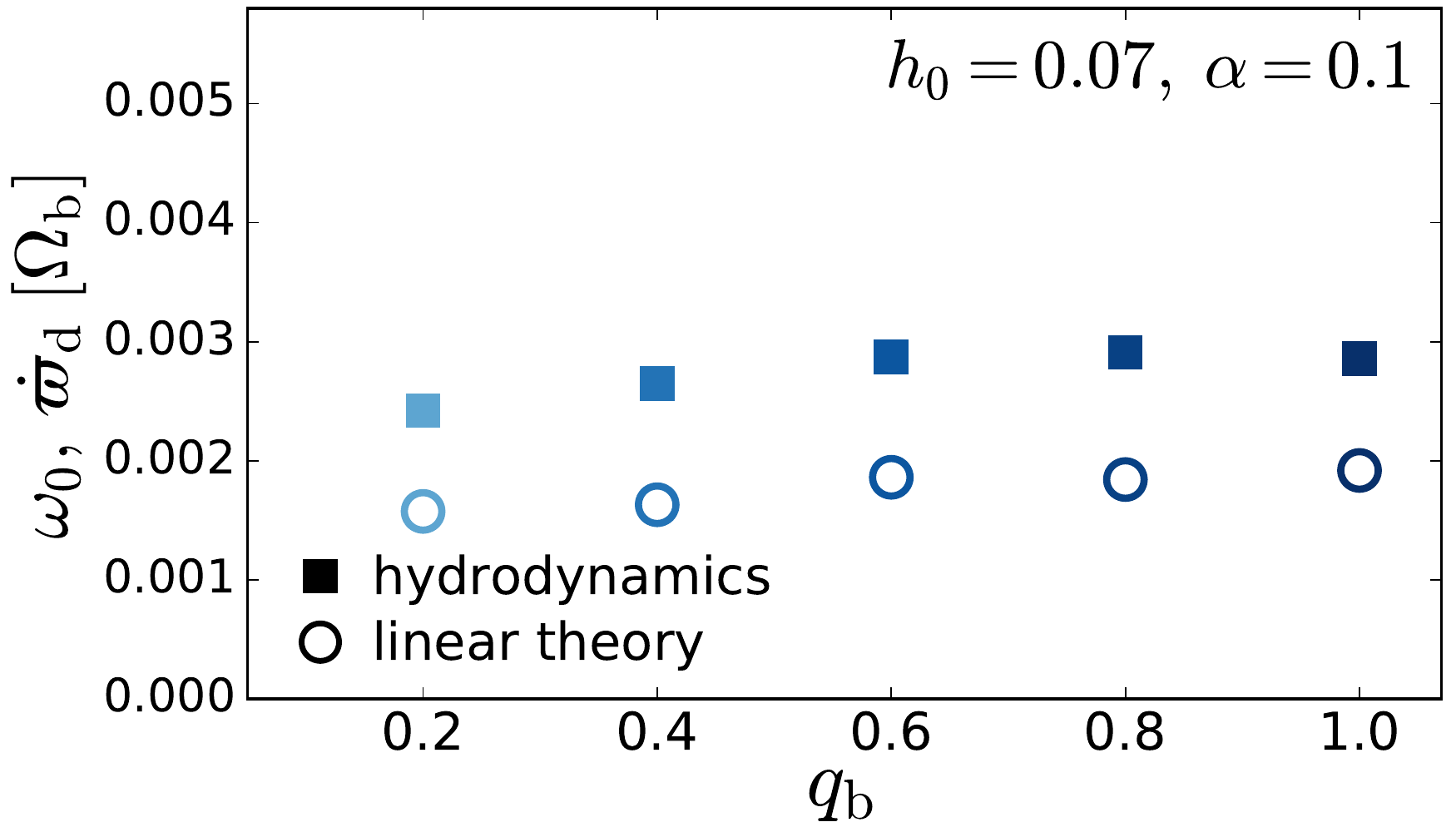}
\caption{Linear fundamental-mode eigenfrequencies $\omega_0$  (circles) corresponding to the eigenfunctions of Figure~\ref{fig:comparison_eigenfunction},
compared to empirically measured precession frequencies $\dot{\varomega}_{\rm d}$ (squares) as given by Equation~(\ref{eq:disk_precession}).
The order of the panels is the same as in Figure~\ref{fig:comparison_eigenfunction}.
\label{fig:comparison_eigenfrequencies}}
\end{figure}

%

\section{Discussion}
An exhaustive understanding of how disk and binary eccentricities couple to each other is still lacking.
In this work, we have taken a first step toward a full picture of disk eccentricity by showing
 that CBDs naturally tend to trap eccentric modes in the vicinity of the central cavity. The turning
 points that allow for the trapping of waves are a consequence of the steeply but smoothly truncated CBD profile, 
 which makes our results independent of the boundary conditions.

\subsection{Importance of eccentric disks}
The inevitability of freely precessing disks around binaries can introduce an important change into our understanding of
the physical processes in circumbinary environments. Long-lived disk eccentricities can
(1) determine the dynamical coupling of  binaries and disks;
(2) modify the physical processes within protoplanetary disks, such as circumbinary planet formation, and (3) potentially leave observable imprints in
 the gas kinematics and disk morphologies.
 
A theoretical understanding of CBD eccentricity profiles might aid the interpretation
of observational data. Recently, the ``binary interpretation'' of transitional disks \citep{ire08} 
has attracted renewed interest
as a result of the rich morphological and kinematic signatures
observed in astrophysical disks \citep[see., e.g.,][for the case of HD142527]{pri18b}. Self-consistent eccentricity
profiles will prove powerful tools as a diagnostic for hidden binaries and
disk properties. 

Moreover, eccentricity profiles inferred from observations can provide independent information on a disk's
density and temperature profiles.
For example, Equation~(\ref{eq:approximate_eigenfunction}) is derived assuming
that far from the cavity $\Sigma\propto R^{-1/2}$  (in turn a consequence of $\nu\propto R^{1/2}$)
and that $c_{\rm s}^2\propto R^{-1}$, but a different set of assumptions would result in different eccentricity profiles.

\subsection{Three-dimensional Effects}
In this work, we have focused on gas eccentric modes in two dimensions, since most long-term hydrodynamical simulations
are 2D. However,  in 3D, the RHS Equation~(\ref{eq:eccentricity_equation}) includes the
additional term  $\tfrac{3}{2}{\rm i}\Sigma \tfrac{d}{dR}(c_s^2R^2)E$ \citep{ogi08,tey16}, which depends on $\Sigma$ but not on its derivative, and thus is always
positive. However, the 2D pressure terms in the vicinity of $R_{\rm cav}$
--where the trapped mode lives-- are already positive, thanks to the steep gradients in $\Sigma$. Thus,
 the 3D term can increase the eigenfrequency, but not change its sign. The 
role of the additional term is made clear
by transforming the 3D eccentricity equation into normal form, repeating the procedure of Section~\ref{sec:effective_potential}.
The effective potential now becomes
\begin{equation}\label{eq:effective_potential_3d}
\begin{split}
\omega_{\rm pot}^{\rm (3D)}
&=\omega_{\cal Q}\left(\frac{R}{R_{\rm cav}}\right)^{-\frac{7}{2}}+\\
&~~~~ +\frac{\omega_{\cal P}}{2}\left({\frac{R}{R_{\rm cav}}}\right)^{\!\!\!-\tfrac{3}{2}}
\!\!\!\bigg[\frac{R\Sigma'}{2\Sigma}+ \left(\!\frac{R\Sigma'}{2\Sigma}\!\right)^2 - \frac{R^2\Sigma''}{2\Sigma}+\frac{3}{4} \bigg]
\end{split}
\end{equation}
Remarkably, the only difference between $\omega_{\rm pot}$ (Equation~\ref{eq:effective_potential}) 
and $\omega_{\rm pot}^{\rm (3D)}$ (Equation~\ref{eq:effective_potential_3d}) is the numerical factor
of $\pm\tfrac{3}{4}$ inside the square brackets, which is much smaller than the $\Sigma$-dependent terms
near the peak of the effective potential (Figure~\ref{fig:wave_diagram}, left panel). Consequently,
eccentric modes are essentially unaffected by 3D terms provided that they are confined by turning points.
Note that the radial coordinate in $\omega_{\rm pot}^{\rm (3D)}$ scales with $R_{\rm cav}$, which implies
that larger cavities would simply displace the mode further out and slow down the precession rate, but not change the sign of $\omega_0$.
The robustness of eccentric mode trapping might explain why \citet{moo19} found no major differences
between 2D and 3D (coplanar) simulations of circumbinary accretion.

\subsection{Future work: eccentric binaries}
In this work, we have limited ourselves to the study of circular binaries. Evidently, a comprehensive study 
of binary disk interaction must include eccentric binaries, since these binaries are known to exist
in the T Tauri phase of stellar evolution \citep[e.g.][]{tof17a,tof19}. The gas dynamics around moderate-to-high eccentricity binaries 
is rich and complex, and can differ substantially from their circular counterparts \citep{mun16b}. Eccentric binaries
accreting from VSS disks might see their eccentricity damped or excited \citep{mun19}, suggesting that there may exist an equilibrium
eccentricity at $e_{\rm b}\sim0.2-0.4$ \citep[see also][]{roe11}, in analogy to a similar phenomenon observed for migrating gas giants \citep{duf15}.
Precessing eccentric disks around binaries may be the culprits
of the  ``alternating preferential accretion'' phenomenon \citep{dun15,mun16b}, which
could explain why primary accretion is dominant in the eccentric T Tauri  binary TWA~3A \citet{tof19},
despite most simulations suggesting that preferential accretion is invariably onto the secondary \citep[e.g.,][]{bat00a,far14,mun20a}.
 
For finite binary eccentricity $e_{\rm b}$, the general solution
$E$ will thus consist of a complementary (homogeneous) solution and a particular one, i,e.,  a ``free'' mode accompanied by a ``forced'' mode.
 In principle, for large enough forced eccentricity, the disk's eccentricity vector will rotate around the tip of the forced eccentricity
rather the origin, i.e., $\varomega_{\rm d}$ will librate instead of circulating. In upcoming work, we will simulate CBDs around eccentric binaries, probing the limitations of linear theory.

\section{Summary and Conclusions}
We have studied the properties of eccentricity modes in accretion disks 
around circular binaries
using linear analysis and direct hydrodynamical simulations. Our findings are:
\begin{itemize}
\item[$(i)$] A linear eigenmode analysis shows that steeply truncated circumbinary disks 
trap eccentricity modes between naturally arising turning points.
Often, only one fundamental mode of low frequency ${\omega_0\sim{\cal O}(10^{-3}\Omega_{\rm b})}$ is allowed. This mode precesses in a prograde way, 
 and $\omega$ closely tracks $\omega_{\cal Q}$ --the test-particle precession rate around a binary-- even when pressure appears to dominate
over the quadrupole.

\item[$(ii)$] The linear analysis shows that, when $q_{\rm b}\lesssim1$, the eccentricity profiles are
concentrated toward the edge of the circumbinary cavity (at a radius of $~2a_{\rm b}$), and
decrease in a tapered power-law fashion, with an exponential drop-off at $\sim10a_{\rm b}$.
For $q_{\rm b}\ll1$ and/or $h_0\gtrsim0.2$, the modes are poorly confined, effectively extending
out to many times the binary separation $a_{\rm b}$.

\item[$(iii)$] We have carried out non-linear hydrodynamical simulations of circumbinary disks in viscous steady state
for  $0.2\leq q_{\rm b}\leq 1$ and different values of $h_0$ and $\alpha$.
These simulations develop a steady-state, coherently precessing eccentricity profile.
The precession rates and the radial dependence of the eccentricity are in good agreement with
our analytical and numerical linear calculations, confirming that simulations around circular binaries
develop free modes sustained by the close balance between excitation and damping.
\end{itemize}


\section*{Acknowledgements}
We thank Wing-Kit Lee for helpful  discussions and
Tomoaki Matsumoto for comments on the manuscript.
 YL acknowledges  NSF grant AST1352369.

\bibliographystyle{apj}
%

%

\vspace{-0.0in}
\appendix
\section{Higher order Eigenmodes}\label{app:a}
%
When $h_0\lesssim0.05$, the quantization condition~(\ref{eq:quantum_condition})
can be satisfied by more than one frequency. This new behavior is illustrated in Figure~\ref{fig:wave_diagram_2},
which shows that the function
$\omega_{\rm pot}$ (in units of $\tfrac{1}{2}h_0^2\Omega_{\rm b}$, left panel) is taller and wider than that shown in Figure~\ref{fig:wave_diagram} for when
$h_0$ is reduced from 0.1 to 0.03. This corresponds to 
changing $\omega_{\cal P}/\omega_{\cal Q}$ from $0.33$ to $0.03$. 
This effective potential now allows for two trapped modes, as highlighted by the closed contours of the DRM  (middle panel). A numerical
solution of the BVP (right panel) shows that a lower frequency one-node mode can accompany the fundamental mode.

\begin{figure*}[hb!]
\includegraphics[width=0.62\textwidth]{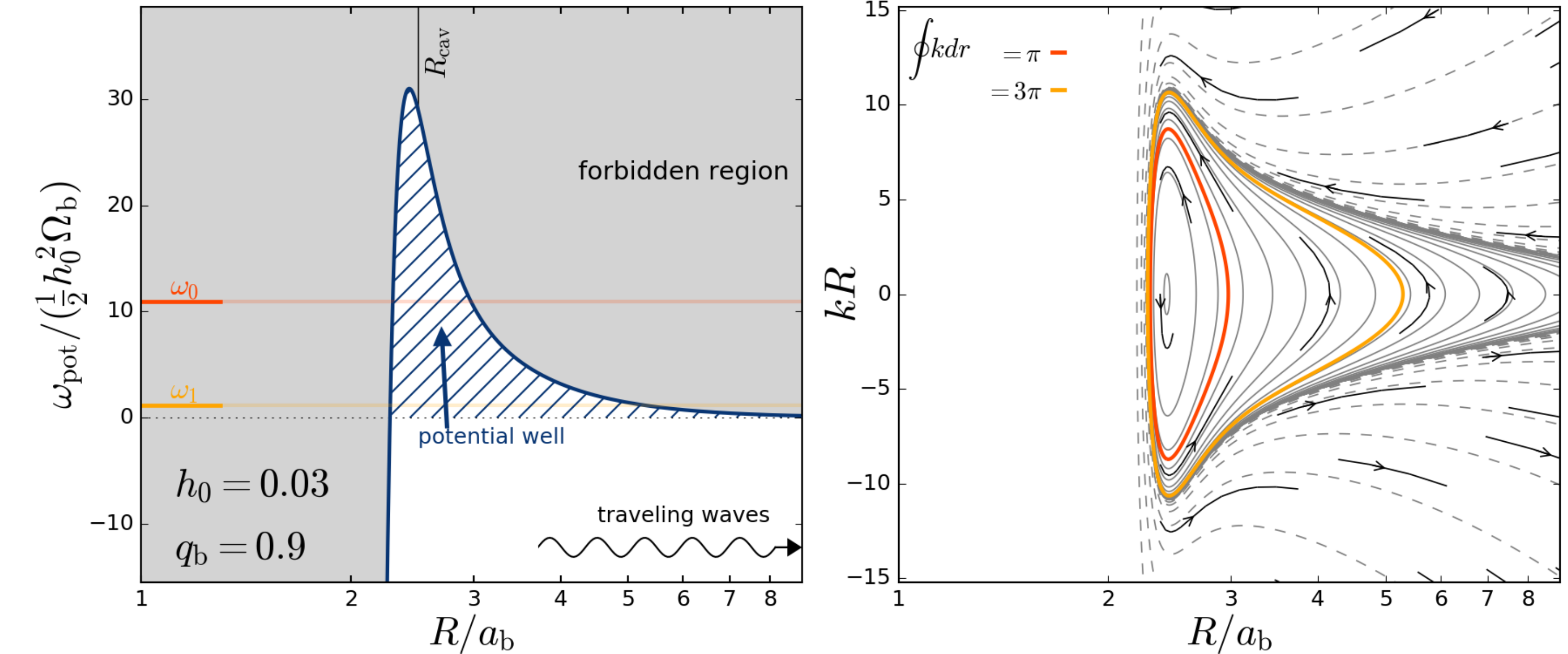}
\includegraphics[width=0.375\textwidth]{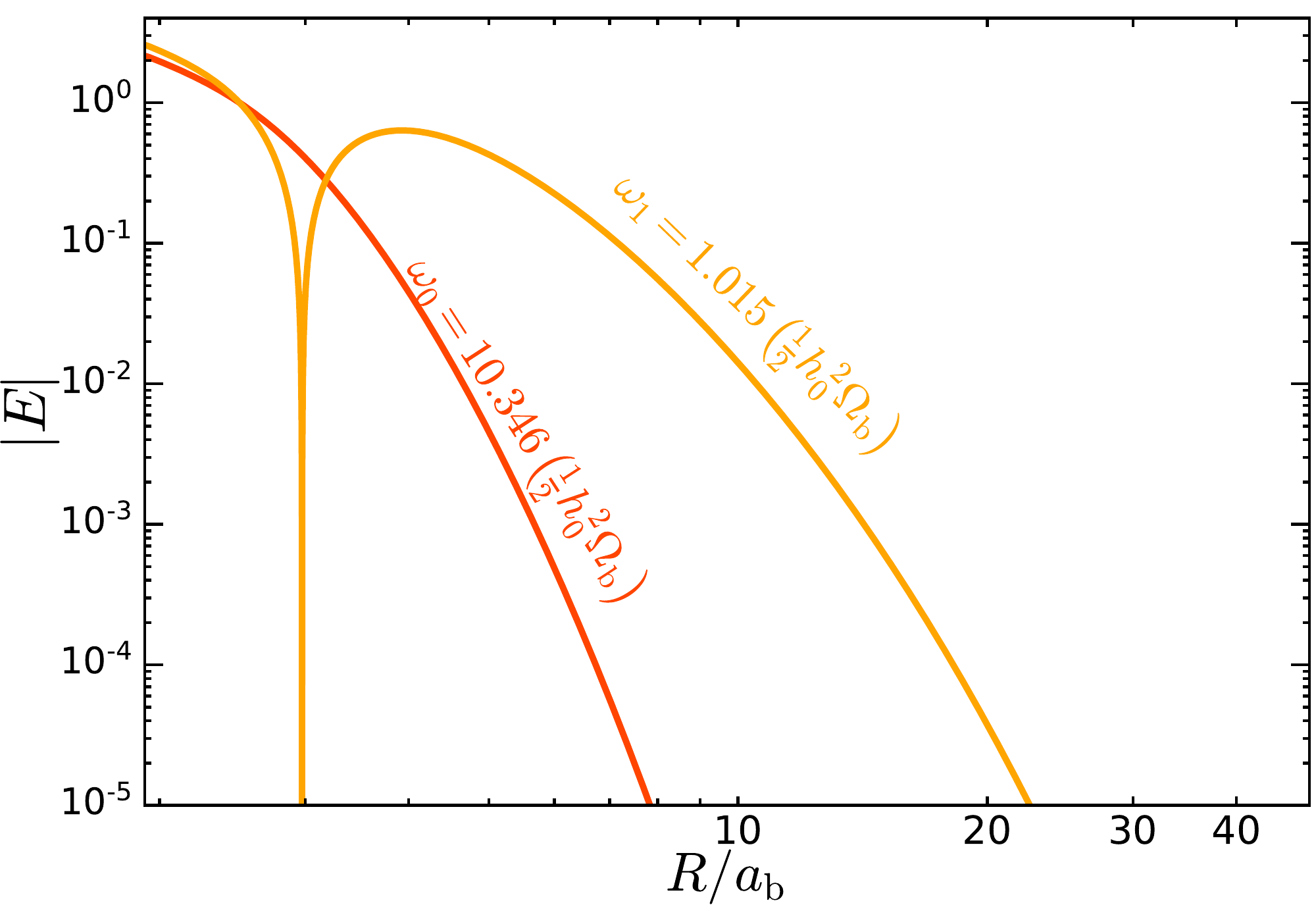}
\caption{Left and middle panels: similar to Figure~\ref{fig:wave_diagram}, but for $h_0=0.03$, which allows
for the existence of  a fundamental mode ($n=0$, red) and a first harmonic ($n=1$, orange) to be trapped.  Right panel: eigenfunctions
associated to the $n=0$ (red) and $n=1$ (orange) modes; the first harmonic contains a node at $R\approx 3a_{\rm b}$
and its frequency $\omega_1$ is 10 times lower than that of the fundamental mode $\omega_0$.
\label{fig:wave_diagram_2}}
\end{figure*}

\section{Quantum Harmonic Oscillator}\label{app:b}
In the vicinity of $\xi=\xi_{\rm peak}$, the term in square brackets in Equation~(\ref{eq:eccentricity_liouville_form}) can be expanded
to quadratic order in $\xi$, simplifying the eccentricity equation to
\begin{equation}\label{eq:eccentricity_liouville_approx}
\frac{9\omega_{\cal P}}{32}\frac{d^2Y}{d\xi^2}+\bigg[
 25\omega_{\cal P}+1.13\omega_{\cal Q}
  +260\omega_{\cal P}(\xi-\xi_{\rm peak})
 -15000\omega_{\cal P}(\xi-\xi_{\rm peak})^2
 \bigg]Y=\omega Y~,
\end{equation}
which is of the generic form
\begin{equation}\label{eq:quadratic_potential}
\hat{H}Y\equiv
\left[\frac{A}{2}\frac{d^2}{d\xi^2}
+B +C(\xi-\xi_0)-\frac{D}{2}(\xi-\xi_0)^2
\right]\psi=\omega Y~.
\end{equation}
If we transform the independent coordinate to $x=\left({D}/{A}\right)^{1/4}\left(\xi-\xi_0-\frac{C}{D}\right)$,
Equation~(\ref{eq:quadratic_potential}) can be cast into a quantum harmonic oscillator (QHO) form 
\begin{equation}
\left[-\frac{1}{2}\frac{d^2}{dx^2}
+\frac{x^2}{2}
\right]Y=\left(AD\right)^{-1/2}\left[\frac{C^2}{2D}+B-\omega\right]Y\equiv \tilde{\omega}Y~,
\end{equation}
which has exact eigenfrequencies $\tilde{\omega}_n=n+1/2$, and thus the operator $\hat{H}$ in  Equation~(\ref{eq:quadratic_potential})  has eigenvalues
\begin{equation}
\omega_n^{\rm QHO} = - \sqrt{AD}\left(n+\frac{1}{2}\right)+\frac{C^2}{2D}+B~~.
\end{equation}
Since $A=(9/16)\omega_{\cal P}$, $B\approx 25\omega_{\cal P}+1.13\omega_{\cal Q}$, $C\approx260\omega_{\cal P}$ and $D=3\times10^4\omega_{\cal P}$,
the eccentricity eigenvalues can be approximated by
\begin{equation}
\omega_n^{\rm QHO}=1.13\omega_{\cal Q}+26.1\omega_{\cal P}-130\omega_{\cal P}\left(n+\frac{1}{2}\right)~.
\end{equation}
%


\section{Approximated Quantization Condition}\label{app:c}
The quantization condition (\ref{eq:quantum_condition}) between two
turning points, $R_{\rm tp,1}$ (left) and $R_{\rm tp,2}$ (right),  cannot be solved analytically for $\omega_0$. However,
an approximate expression can be obtained if we make three assumptions: $(i)$ that the modes are shallow (i.e., that $\omega_0\ll\max[\omega_{\rm pot}]$),
$(ii)$  that $R_{\rm tp,1}$ is the same for all eigenmodes due to the steep decline of $\omega_{\rm pot}$, and $(iii)$
 that, at the right turning point where $\omega_0=\omega_{\rm pot}$, the disk is approximately a power-law and $\omega_{\rm pot}\approx\Omega f_0$.
Then, we approximate Equation~(\ref{eq:quantum_condition}) with 
 \begin{equation}\label{eq:quantum_condition_approx}
 (kR)_{\rm max}\Delta \ln R \sim \pi
 \end{equation}
where $\Delta\ln R\simeq \ln (R_{\rm tp,2}/  R_{\rm tp,1})$, and where assumption $(iii)$ implies that 
the right turning point satisfies $R_{\rm tp,2}\approx R_{\rm cav} (\omega_0/\omega_{\cal Q})^{-2/7}$. Since the modes are shallow,
$kR\simeq (R/R_{\rm cav})^{3/4}({2\omega_{\rm pot}/\omega_{\cal P}})^{1/2}$ (Equation~\ref{eq:eccentricity_scaled_form}) and
thus, in order to solve  Equation~(\ref{eq:quantum_condition_approx}) for $\omega_0$, we just need
to know $\max[(R/R_{\rm cav})^{3/4}({2\omega_{\rm pot}/\omega_{\cal P}})^{1/2}]$. If $C_{\cal P}\omega_{\cal P}$ is the local maximum
of the pressure contribution to $\omega_{\rm pot}$, then we can rearrange the effective potential as
$\omega_{\rm pot}= C_{\cal P}\omega_{\cal P}\left[V_{\rm press}(x)+\epsilon x^{-7/2}\right]$,
where dimensionless function $V_{\rm press}$ evaluates to 1 at $x\equiv R/R_{\rm cav}\simeq 1$,
and where we have defined $\epsilon\equiv {\omega_{\cal Q}}/({C_{\cal P}\omega_{\cal P}})$ which is small
for all of the simulations carried out in this work. Thus, if we replace
$(kR)_{\rm max}$ with $\sqrt{2C_{\cal P}}(1+{\epsilon}/{2})$
and $\Delta\ln R$ with $\ln [(R_{\rm cav}/ R_{\rm tp,1}) (\omega_0/\omega_{\cal Q})^{-2/7}]$
in Equation~(\ref{eq:quantum_condition_approx}), then we can solve for $\omega_0$. We empirically find,
for our assumed $\Sigma$ profile, that   $2C_{\cal P}\simeq50$ and $R_{{\rm tp},1}\simeq0.92 R_{\rm cav}$,
and therefore we have
\begin{equation}
\omega_0
\sim {\rm e}^{-1.28+\pi\tfrac{\epsilon}{4}}~
\omega_{\cal Q}~.
 \end{equation}
Therefore, the mode tapering radius (Equation~\ref{eq:tapering_radius}) is, to first order in $\epsilon$,
\begin{equation}
\lambda=R_{\rm cav}\left(\frac{9}{32}\frac{\omega_{\cal P}}{\omega_0}\right)^{2/3}
\sim R_{\rm cav}\left(\frac{\omega_{\cal P}}{\omega_{\cal Q}}\right)^{2/3}
\left[1-0.02\frac{\omega_{\cal Q}}{\omega_{\cal P}}\right]~.
 \end{equation}

\end{document}